\documentclass[aps,prd,twocolumn,superscriptaddress,preprintnumbers]{revtex4}
\usepackage{graphicx}

\def\cB{{\cal{B}}}
\def \s{\sqrt{2}}
\def \st{\sqrt{3}}
\def \sx{\sqrt{6}}

\def\be{\begin{equation}}
\def\ee{\end{equation}}
\def\bea{\begin{eqnarray}}
\def\eea{\end{eqnarray}}
\def\bean{\begin{eqnarray*}}
\def\eean{\end{eqnarray*}}
\def\bary{\begin{array}}
\def\eary{\end{array}}
\def\half{\frac{1}{2}}

\def\mco{\multicolumn}

\def\lbar{\overline}
\def \bl{\bar \lambda}
\def \ob{\overline{B}^0}

\def\ubar{{\bar u}}
\def\cbar{{\bar c}}

\def\dbar{{\bar d}}
\def\sbar{{\bar s}}
\def\bbar{{\bar b}}
\def\bit{\begin{itemize}}
\def\eit{\end{itemize}}
\def\ol{\overline}

\begin{document}

\preprint{\vbox{ \hbox{ANL-HEP-PR-03-061}
                 \hbox{EFI-03-36}
                 \hbox{hep-ph/0307395}
         }}
\title{CHARMLESS $B \to VP$ DECAYS USING FLAVOR SU(3) SYMMETRY
\footnote{To be submitted to Phys.\ Rev.\ D.}}

\author{Cheng-Wei Chiang}
\email[e-mail: ]{chengwei@hep.uchicago.edu}
\affiliation{Enrico Fermi Institute and Department of Physics,
University of Chicago, 5640 S. Ellis Avenue, Chicago, IL 60637}
\affiliation{HEP Division, Argonne National Laboratory,
9700 S. Cass Avenue, Argonne, IL 60439}
\author{Michael Gronau}
\email[e-mail: ]{gronau@physics.technion.ac.il}
\affiliation{Enrico Fermi Institute and Department of Physics, 
University of Chicago, 5640 S. Ellis Avenue, Chicago, IL 60637}
\affiliation{Department of Physics, Technion -- Israel Institute of
Technology, Haifa 32000, Israel (Permanent Address)}
\author{Zumin Luo}
\email[e-mail: ]{zuminluo@midway.uchicago.edu}
\affiliation{Enrico Fermi Institute and Department of Physics, 
University of Chicago, 5640 S. Ellis Avenue, Chicago, IL 60637}
\author{Jonathan L.~Rosner}
\email[e-mail: ]{rosner@hep.uchicago.edu}
\affiliation{Enrico Fermi Institute and Department of Physics, 
University of Chicago, 5640 S. Ellis Avenue, Chicago, IL 60637}
\author{Denis A.~Suprun}
\email[e-mail: ]{d-suprun@uchicago.edu}
\affiliation{Enrico Fermi Institute and Department of Physics, 
University of Chicago, 5640 S. Ellis Avenue, Chicago, IL 60637}

\date{\today}

\begin{abstract}
The decays of $B$ mesons to a charmless vector ($V$) and pseudoscalar ($P$)
meson are analyzed within a framework of flavor SU(3) in which symmetry
breaking is taken into account through ratios of decay constants in
tree ($T$) amplitudes.  The magnitudes and relative phases of tree and
penguin amplitudes are extracted from data; the symmetry assumption is tested;
and predictions are made for rates and $CP$ asymmetries in as-yet-unseen
decay modes.  A key assumption for which we perform some tests and suggest
others is a relation between penguin amplitudes in which the spectator quark
is incorporated into either a pseudoscalar meson or a vector meson.
Values of $\gamma$ slightly restricting the range currently allowed by fits
to other data are favored, but outside this range there remain acceptable
solutions which cannot be excluded solely on the basis of present $B \to VP$
experiments.

\end{abstract}


\maketitle

\section{INTRODUCTION \label{sec:intro}}

$B$ meson decays are a rich source of information on fundamental phases of weak
charge-changing couplings, as encoded in the Cabibbo-Kobayashi-Maskawa (CKM)
matrix.  Decays to charmless final states, many of which occur at branching
ratios exceeding $10^{-5}$, are particularly useful, since many of them involve
more than one significant quark subprocess and thus have the potential for
displaying direct $CP$ asymmetries. To interpret such data one must disentangle
information on CKM (weak) phases from strong-interaction final-state phases.

In $B \to PP$ decays, where $P$ is a charmless pseudoscalar meson,
flavor SU(3) symmetries have been employed \cite{DZ,SW,Chau,%
Gronau:1994rj,Gronau:1995hn} to extract weak phases in such decays as $B^0 \to
\pi^+ \pi^-$ and various charge states of $B \to K \pi$ (see, e.g., the recent
reviews of \cite{conf} and a recent analysis \cite{comb03} of $B \to K \pi$).
The decays $B \to VP$, where $V$ is a charmless vector meson, involve
more invariant amplitudes, since one cannot use Bose statistics to simplify
the decays, in contrast to the case of two spinless final pseudoscalars
in the same meson multiplet \cite{DZ,SW,Chau,Dighe:1998wj}.  Nonetheless, after
the first report of a charmless $B \to VP$ decay \cite{Bergfeld:1998ik}, it
became possible to perform such analyses by using rates and $CP$ asymmetries in
some decays to predict others \cite{Dighe:1998wj,VPUP,Gronau:2000az,%
Chiang:2001ir}.

In the present paper, following upon our recent analysis of $B \to PP$ decays
\cite{Chiang:2003rb}, we analyze $B \to VP$ decays within flavor SU(3),
incorporating symmetry breaking through ratios of meson decay
constants in tree ($T$) amplitudes.  The magnitudes and relative phases of
invariant amplitudes are extracted from data; the symmetry assumption is
tested; and predictions are made for rates and $CP$ asymmetries in
as-yet-unseen decay modes.

Our approach differs from ones involving {\it a priori}
calculations of $B \to V P$ decay rates and $CP$ asymmetries involving QCD and
factorization.  Factorization was applied to these decays in Refs.\ 
\cite{Kramer:1995,Deshpande:1997rr,Ali:1998eb,Ali:1998gb}.  The QCD
factorization methods of Refs.\ \cite{BBNS,Beneke:2002jn} were considered for
$B \to VP$ decays in Refs.\ \cite{Beneke:2002jn,Yang:2000,Du:2002,Sun:2002,%
Guo:2003wb,Beneke:2003}.  Many of these authors were able to fit some data but
could not reproduce those processes dominated by strangeness-changing penguin
amplitudes, which others have argued should be enhanced \cite{charming,%
Aleksan,Isola:2003,Groot:2003}.  Our method, by contrast,
relies on assumptions of isospin and SU(3) flavor symmetry, provides tests
of these assumptions, and is capable of extracting strong final-state
phases from data rather than needing to predict them.  It is similar
to the analysis of $B \to \rho^\pm \pi^\mp$ and $B \to \rho^\mp K^\pm$
in Ref.\ \cite{Hocker} and of $B \to K^{*\pm} \pi^\mp$ in Ref.\ \cite{Sun}
(which uses extensive data on $B \to \rho^\pm \pi^\mp$), but we are concerned
with a wider set of $B \to VP$ decays.

The present analysis has considerable sensitivity to the CKM phase
$\gamma$. This is driven in part by the pattern of tree-penguin
interference in a wide variety of $B \to VP$ decays, and in part by the
incorporation of time-dependent information on $B \to \rho \pi$, as has
also been noted in Ref.\ \cite{Beneke:2003}.

We review notation and conventions for amplitudes in Section \ref{sec:not}.  We
average currently known experimental rates and $CP$ asymmetries from the CLEO,
BaBar, and Belle Collaborations and use these averages to obtain magnitudes of 
amplitudes in Section \ref{sec:amp}.  We then show how to extract invariant
amplitudes (identified with specific flavor topologies) in Section
\ref{sec:extract} by fitting the experimental amplitudes and $CP$ asymmetries.
The simplest fit assumes a relation between penguin amplitudes \cite{HJLP} in
which the spectator quark is incorporated into either a pseudoscalar meson or a
vector meson.  We suggest specific tests of this assumption in Section
\ref{sec:test}.  It is relaxed in Section \ref{sec:var} to see if the
quality of the overall fit improves, and predictions for rates and $CP$
asymmetries for observed and as-yet-unseen $B \to VP$ modes are discussed.
Relations among amplitudes based on the U-spin subgroup of SU(3) are presented
in Section \ref{sec:uspin}, while Section \ref{sec:concl} concludes.
An Appendix discusses $B \to \rho^\mp \pi^\pm$ rates and asymmetries.

\section{NOTATION \label{sec:not}}

We use the following quark content and phase conventions:
\begin{itemize}
\item{ {\it Bottom mesons}: $B^0=d\bbar$, ${\lbar B}^0=b\dbar$, $B^+=u\bbar$,
    $B^-=-b\ubar$, $B_s=s\bbar$, ${\lbar B}_s=b\sbar$;}
\item{ {\it Charmed mesons}: $D^0=-c\ubar$, ${\lbar D}^0=u\cbar$, $D^+=c\dbar$,
    $D^-=d\cbar$, $D_s^+=c\sbar$, $D_s^-=s\cbar$;}
\item{ {\it Pseudoscalar mesons}: $\pi^+=u\dbar$,
    $\pi^0=(d\dbar-u\ubar)/\sqrt{2}$, $\pi^-=-d\ubar$, $K^+=u\sbar$,
    $K^0=d\sbar$, ${\lbar K}^0=s\dbar$, $K^-=-s\ubar$,
    $\eta=(s\sbar-u\ubar-d\dbar)/\sqrt{3}$,
    $\eta^{\prime}=(u\ubar+d\dbar+2s\sbar)/\sqrt{6}$;}
\item{ {\it Vector mesons}: $\rho^+=u\dbar$, $\rho^0=(d\dbar-u\ubar)/\sqrt{2}$,
    $\rho^-=-d\ubar$, $\omega=(u\ubar+d\dbar)/\sqrt{2}$, $K^{*+}=u\sbar$,
    $K^{*0}=d\sbar$, ${\lbar K}^{*0}=s\dbar$, $K^{*-}=-s\ubar$, $\phi=s\sbar$.}
\end{itemize}

In the present approximation there are four types of independent amplitudes: a
``tree'' contribution $t$; a ``color-suppressed'' contribution $c$; a
``penguin'' contribution $p$; and a ``singlet penguin'' contribution $s$, in
which a color-singlet $q \bar q$ pair produced by two or more gluons or by a
$Z$ or $\gamma$ forms an SU(3) singlet state.  We neglect smaller contributions
from an ``exchange'' amplitude $e$, an ``annihilation'' amplitude $a$, and a
``penguin annihilation'' amplitude $pa$.  The amplitudes we retain contain both
the leading-order and electroweak penguin contributions:
\be\bary{lll}
\label{eq:dict}
t \equiv T + P_{\rm EW}^C ~, &\quad& c \equiv C + P_{\rm EW} ~, \\
p \equiv P - \frac{1}{3} P_{\rm EW}^C ~, &\quad&
s \equiv S - \frac{1}{3} P_{\rm EW} ~,
\eary\ee
where the capital letters denote the leading-order contributions
(\cite{Gronau:1994rj,Gronau:1995hn,Dighe:1995gq,DGReta}) while $P_{\rm EW}$ and
$P_{\rm EW}^C$ are respectively color-favored and color-suppressed electroweak
penguin (EWP) amplitudes \cite{Gronau:1995hn}.  We shall denote $\Delta S = 0$
transitions by unprimed quantities and $|\Delta S| = 1$ transitions by primed
quantities.  For $VP$ decay modes, the subscript $V$ or $P$ denotes the
final-state meson (vector or pseudoscalar) incorporating the spectator quark.
Thus, for example, a color-favored $\Delta S = 0$ tree amplitude in which the
spectator quark is incorporated into a pseudoscalar meson will be 
denoted $t_P$.
Although some $B \to VV$ decay processes have been seen, we shall not discuss
them further here.

For the $\bbar \to \dbar$ and $\bbar \to \ubar u \dbar$ transitions, an
educated guess of the hierarchies among the amplitudes is given in
Ref.~\cite{Gronau:1995hn,Chiang:2001ir,Chiang:2003rb}.  For
$|\Delta S| = 1$ transitions, $c'$ contains an electroweak penguin amplitude at
the next order.  Therefore, we put $c'$ together with $t'$ at the same order.
Similarly, since part of the singlet amplitude is the electroweak penguin, $s'$
is at least of order $P'_{\rm EW}$.

The partial decay width of two-body $B$ decays is
\be
\label{eq:width}
\Gamma(B \to M_1 M_2)
= \frac{p_c}{8 \pi m_B^2} |{\cal A}(B \to M_1 M_2)|^2 ~,
\ee
where $p_c$ is the momentum of the final state meson in the rest frame of $B$,
$m_B$ is the $B$ meson mass, and $M_1$ and $M_2$ can be either pseudoscalar or
vector mesons.  Using Eq.~(\ref{eq:width}), one can extract the magnitude of
the invariant amplitude of each decay mode from its experimentally measured
branching ratio.  To relate partial widths to branching ratios, we use the
world-average lifetimes $\tau^+ = (1.653 \pm 0.014)$ ps and $\tau^0 = (1.534
\pm 0.013)$ ps computed by the LEPBOSC group \cite{LEPBOSC}.  Unless otherwise
indicated, for each branching ratio quoted we imply the average of a process
and its $CP$-conjugate.

Two phase conventions are in current use for the penguin amplitudes,
depending on whether one considers them to be dominated by the CKM factors
$V^*_{tb} V_{tq}$ ($q = s,d$), or integrates out the $t$ quark, uses the
unitarity relation $V^*_{tb} V_{tq} = - V^*_{cb} V_{cq} - V^*_{ub} V_{uq}$,
and absorbs the $V^*_{ub} V_{uq}$ term into a redefined tree amplitude.
Here we adopt the latter convention.  For a discussion of the relation
between the two see, e.g., Ref.\ \cite{Gronau:2002gj}.  Thus both the
strangeness-changing and strangeness-preserving penguin amplitudes will be
taken to have real weak phases in this discussion.

\section{EXPERIMENTAL DATA AND AMPLITUDE DECOMPOSITIONS \label{sec:amp}}

The experimental branching ratios and $CP$ asymmetries from the CLEO, BaBar, 
and
Belle Collaborations are summarized and averaged in Tables \ref{tab:dS0data}
(for $\Delta S = 0$ transitions) and \ref{tab:dS1data} (for $|\Delta S| =1$
transitions).  Data are current up to and including the 2003 Lepton-Photon
Symposium at Fermilab.  We use the Particle Data Group method \cite{PDG} for
performing averages, including a scale factor $S \equiv [\chi^2/(N-1)]^{1/2}$
when the $\chi^2$ for an average of $N$ data points exceeds $N-1$.  (The
Heavy Flavor Averaging Group \cite{HFAG} does not use this scale factor.  In
other respects our averages agree with theirs when inputs are the same.)
The corresponding experimental amplitudes, extracted from
partial decay rates using Eq.\ (\ref{eq:width}), are shown in Tables
\ref{tab:dS0} and \ref{tab:dS1}.  In these tables we also give the theoretical
expressions for these amplitudes (see also Refs.\ \cite{Dighe:1998wj,VPUP,%
Chiang:2001ir}) and, anticipating the results of the next section, the
magnitudes of contributions to the observed amplitude of the invariant
amplitudes $|\cal{T}|$ and $|\cal{P}|$ or $|\cal{T}'|$ 
and $|\cal{P}'|$ in one fit to the data.  These contributions
include Clebsch-Gordan coefficients.  $CP$ asymmetries are defined as
\be
A_{CP}(B \to f)\equiv \frac{\Gamma(\bar B \to \bar f)
- \Gamma(B \to  f)}{\Gamma(\bar B \to \bar f) + \Gamma(B \to  f)}~.
\ee
By comparing the magnitudes of individual contributions with experimental
amplitudes, one can tell whether one contribution dominates or whether
constructive or destructive interference between two contributions is
favored.

\begin{table*}[t]
\caption{Experimental branching ratios of selected $\Delta S = 0$ decays of $B$
  mesons.  $CP$-averaged branching ratios are quoted in units of $10^{-6}$.
  Numbers in parentheses are upper bounds at 90 \% c.l.  References are given
  in square brackets.  Additional line, if any, gives the $CP$ asymmetry
  $A_{CP}$.  The error in the average includes the scale factor $S$ when this
  number is shown in parentheses.
\label{tab:dS0data}}
\begin{ruledtabular}
\begin{tabular}{llllll}
 & Mode & CLEO & BaBar & Belle & Avg. \\ 
\hline
$B^+ \to$
    & $\lbar{K}^{*0} K^+$ 
        & $0.0^{+1.3+0.6}_{-0.0-0.0} \; (<5.3)$ \cite{Jessop:2000bv}
        & -
        & -
        & $<5.3$ \\
    & $\rho^0 \pi^+$ 
        & $10.4^{+3.3}_{-3.4}\pm2.1$ \cite{Jessop:2000bv}
        & $9.3 \pm 1.0 \pm 0.8$ \cite{Ocariz:2003}
        & $8.0^{+2.3}_{-2.0}\pm0.7$ \cite{Gordon:2002yt}
        & $9.1 \pm 1.1$ \\
    &   & -
        & $-0.17 \pm 0.11 \pm 0.02$ \cite{Ocariz:2003} 
        & -
        & $-0.17 \pm 0.11$ \\
    & $\rho^+ \pi^0$ 
        & $<43$ \cite{Jessop:2000bv}
        & $11.0 \pm 1.9 \pm 1.9$ \cite{Ocariz:2003} 
        & -
        & $11.0 \pm 2.7$ \\
    &   & -
        & $0.23 \pm 0.16 \pm 0.06$ \cite{Ocariz:2003} 
        & -
        & $0.23 \pm 0.17$ \\
    & $\rho^+ \eta$ 
        & $4.8^{+5.2}_{-3.8} \; (<15)$ \cite{Richichi:1999kj}
        & $10.5^{+3.1}_{-2.8} \pm 1.3$ \cite{Aubert:2003ru}
        & $<6.2$ \cite{Aihara2003}
        & $8.9 \pm 2.7$  \\
    &   & -
        & $0.06 \pm 0.29 \pm 0.02$ \cite{Aubert:2003ru}
        & -
        & $0.06 \pm 0.29$ \\
    & $\rho^+ \eta'$ 
        & $11.2^{+11.9}_{-7.0} \; (<33)$ \cite{Richichi:1999kj}
        & $14.0^{+5.1}_{-4.6} \pm 1.9$ \cite{Aubert:2003ru} 
        & -
        & $13.3 \pm 4.5$  \\
    & $\omega \pi^+$ 
        & $11.3^{+3.3}_{-2.9}\pm1.4$ \cite{Jessop:2000bv}
        & $5.4 \pm 1.0 \pm 0.5$ \cite{Aubert:2003fa}
        &  $5.7^{+1.4}_{-1.3}\pm0.6$ \cite{Abe2003}
        & $5.9 \pm 1.1~(S=1.23)$ \\
    &   & $-0.34 \pm 0.25 \pm 0.02$ \cite{Chen:2000hv}
        & $0.04 \pm 0.17 \pm0.01$ \cite{Aubert:2003fa} 
        & $0.48^{+0.23}_{-0.20} \pm 0.02$ \cite{Abe2003}
        & $0.10 \pm 0.21~(S=1.84)$ \\
    & $\phi \pi^+$
        & $< 5$ \cite{Bergfeld:1998ik}
        & $<0.41$ \cite{Aubert:2003tk}
        & -
        & $<0.41$ \\
\hline
$B^0 \to$
    & $\rho^{\mp} \pi^{\pm}$ 
        & $27.6^{+8.4}_{-7.4}\pm4.2$ \cite{Jessop:2000bv}
        & $22.6 \pm 1.8 \pm 2.2$ \cite{Aubert:2003wr}
        & $29.1^{+5.0}_{-4.9} \pm 4.0$ \cite{Abe:2003rj}
        & $24.0 \pm 2.5$ \\
    &   & - & $-0.11 \pm 0.06 \pm 0.03$ \cite{Jawahery:2003} 
        & $-0.38^{+0.19+0.04}_{-0.21-0.05}$ \cite{Abe:2003rj}
        & $-0.14 \pm 0.08~(S=1.31)$ \\
    & \mco{1}{r}{$\rho^- \pi^+$}
        & -
    & $9.5 \pm 2.0^{~a}$
        & -
        & $10.2 \pm 2.0^{~b}$ \\
    & 
        & -
        & $-0.52^{+0.17}_{-0.19} \pm 0.07$ \cite{Jawahery:2003}
        & -
        & $-0.54 \pm 0.19^{~b}$ \\
    & \mco{1}{r}{$\rho^+ \pi^-$}
        & -
        & $13.1 \pm 2.3^{~a}$
        & -
        & $13.8 \pm 2.2^{~b}$ \\
    & 
        & -
        & $-0.18 \pm 0.13 \pm 0.05$ \cite{Jawahery:2003}
        & -
        & $-0.16 \pm 0.15^{~b}$ \\
    & $\rho^0 \pi^0$
        & $1.6^{+2.0}_{-1.4}\pm0.8 \; (<5.5)$ \cite{Jessop:2000bv}
        & $0.9 \pm 0.7 \pm 0.5 \; (<2.5)$ 
                                            \cite{Ocariz:2003}
        & $6.0^{+2.9}_{-2.3} \pm 1.2$ \cite{Abe:2003rj}
        & $<2.5$ \\
    & $\rho^0 \eta$ 
        & $2.6^{+3.2}_{-2.6} \; (<10)$ \cite{Richichi:1999kj}
        & - 
        & $<5.5$ \cite{Belle0137}
        & $<5.5$ \\
    & $\rho^0 \eta'$ 
        & $0.0^{+5.8}_{-0.0} \; (<12)$ \cite{Richichi:1999kj}
        & - 
        & $<14$ \cite{Aihara2003}
        & $<12$ \\
    & $\omega \pi^0$ 
        & $0.8^{+1.9+1.0}_{-0.8-0.8} \; (<5.5)$ \cite{Jessop:2000bv} 
        & $-0.3 \pm 1.1 \pm 0.3 \; (<3)$ \cite{Aubert:2001zf}
        & $< 1.9$ \cite{Abe2003} 
        & $< 1.9$ \\
    & $\omega \eta$
        & $< 12$ \cite{Bergfeld:1998ik}
        & -
        & -
        & $< 12$ \\
    & $\omega \eta'$
        & $< 60$ \cite{Bergfeld:1998ik}
        & -
        & -
        & $< 60$ \\
    & $\phi \pi^0$
        & $< 5$ \cite{Bergfeld:1998ik}
        & -
        & -
        & $< 5$ \\
    & $\phi \eta$
        & $< 9$ \cite{Bergfeld:1998ik}
        & -
        & -
        & $< 9$ \\
    & $\phi \eta'$
        & $< 31$ \cite{Bergfeld:1998ik}
        & $< 1.0$ \cite{phietap}
        & -
        & $< 1.0$ \\
\end{tabular}
\end{ruledtabular}
\leftline{$^a$Based on asymmetries quoted in Ref.\ \cite{Jawahery:2003} and
BaBar value of $\cB(B^0 \to \rho^\mp \pi^\pm)$.}
\leftline{$^b$Based on asymmetries quoted in Ref.\ \cite{Jawahery:2003} and
world averages for $\cB$ and $A_{CP}$ for $B^0 \to \rho^\mp \pi^\pm$.}
\end{table*}
%

\begin{table*}
\caption{Same as Table \ref{tab:dS1data} for $|\Delta S| = 1$ decays of $B$
  mesons.
\label{tab:dS1data}}
\begin{ruledtabular}
\begin{tabular}{llllll}
 & Mode & CLEO & BaBar & Belle & Avg. \\
\hline
$B^+ \to$
    & $K^{*0} \pi^+$
        & $7.6^{+3.5}_{-3.0}\pm1.6 \; (<16)$ \cite{Jessop:2000bv}
        & $15.5\pm1.8^{+1.5}_{-4.0}$ \cite{Aubert:2003}
        & $8.5 \pm 0.9^{+0.8+0.8}_{-0.7-0.5}$ \cite{Belle0338}
        & $9.0 \pm 1.4~(S=1.11)$ \\
    & $K^{*+} \pi^0$
        & $7.1^{+11.4}_{-7.1}\pm1.0 \; (<31)$ \cite{Jessop:2000bv}
        & -
        & -
        & $<31$ \\
    & $K^{*+} \eta$
        & $26.4^{+9.6}_{-8.2}\pm3.3$ \cite{Richichi:1999kj}
        & $25.7^{+3.8}_{-3.6} \pm 1.8$ \cite{Aubert:2003ru}
        & $26.5^{+7.8}_{-7.0}\pm3.0$ \cite{Huang:2002ev}
        & $25.9 \pm 3.4$ \\
    &   & -
        & $0.15 \pm 0.14 \pm 0.02$ \cite{Aubert:2003ru}
        & $-0.05^{+0.25}_{-0.30} \pm 0.01$ \cite{Tomura}
        & $0.10 \pm 0.12$ \\
    & $K^{*+} \eta'$
        & $11.1^{+12.7}_{-8.0} \; (<35)$ \cite{Richichi:1999kj}
        & $6.1^{+3.9}_{-3.2} \pm 1.2 \; (< 12)$ \cite{Aubert:2003ru}
        & $<90$ \cite{Aihara2003}
        & $<12$ \\
    & $\rho^0 K^+$
        & $8.4^{+4.0}_{-3.4}\pm1.8 \; (<17)$  \cite{Jessop:2000bv}
        & $3.9 \pm 1.2^{+1.3}_{-3.5} \; (<6.2)$ \cite{Aubert:2003}
        & $3.9 \pm 0.6^{+0.4+0.7}_{-0.3-0.2}$ \cite{Belle0338}
        & $4.1 \pm 0.8$ \\
    & $\rho^+ K^0$
        & $<48$ \cite{Asner:1996hc}
        & -
        & -
        & $<48$ \\
    & $\omega K^+$
        & $3.2^{+2.4}_{-1.9} \pm 0.8 \; (<7.9)$  \cite{Jessop:2000bv}
        & $5.0 \pm 1.0 \pm 0.4$ \cite{Aubert:2003fa}
        & $6.7^{+1.3}_{-1.2} \pm 0.6$ \cite{Abe2003}
       & $5.4 \pm 0.8$  \\
    &   & -
        & $-0.05 \pm 0.16 \pm0.01$ \cite{Aubert:2003fa}
        & $0.06^{+0.20}_{-0.18} \pm 0.01$ \cite{Abe2003}
        & $0.00 \pm 0.12$ \\
    & $\phi K^+$
        & $5.5^{+2.1}_{-1.8} \pm 0.6$ \cite{Briere:2001ue}
        & $10.0^{+0.9}_{-0.8} \pm 0.5$ \cite{Aubert:2003tk}
        & $8.6 \pm 0.8^{+0.6+0.0}_{-0.6-0.3}$ \cite{Belle0338}
        & $9.0 \pm 0.9~(S=1.39)$ \\
    &   & -
        & $0.039 \pm 0.086 \pm0.011$ \cite{Aubert:2003tk}
        & $0.01 \pm 0.12 \pm 0.05$ \cite{Chen:2003}
        & $0.03 \pm 0.07$\\
\hline
$B^0 \to$
    & $K^{*+} \pi^-$
        & $16^{+6}_{-5}\pm2$ \cite{Eckhart:mb}
        & -
        & $14.8^{+4.6+1.5+2.4}_{-4.4-1.0-0.9}$ \cite{Belle0317}
        & $15.3 \pm 3.8$ \\
        &
        & $0.26^{+0.33+0.10}_{-0.34-0.08}$ \cite{Eisenstein:2003yy}
        & -
        & -
        & $0.26 \pm 0.35$ \\
    & $K^{*0} \pi^0$
        & $0.0^{+1.3+0.5}_{-0.0-0.0} \; (<3.6)$ \cite{Jessop:2000bv}
        & -
        & $0.42^{+1.85}_{-1.74} \pm 0.06 \; (<3.5)$ \cite{Belle0317}
        & $0.4 \pm 1.8^{~a}~(<3.5)$ \\
    & $K^{*0} \eta$
        & $13.8^{+5.5}_{-4.6} \pm 1.6$ \cite{Richichi:1999kj}
        & $19.0^{+2.2}_{-2.1} \pm 1.3$ \cite{Aubert:2003ru}
        & $16.5^{+4.6}_{-4.2} \pm 1.2$ \cite{Huang:2002ev}
        & $17.8 \pm 2.0$ \\
    &   & -
        & $0.03 \pm 0.11 \pm 0.02$ \cite{Aubert:2003ru}
        & $0.17^{+0.28}_{-0.25} \pm 0.01$ \cite{Tomura}
        & $0.05 \pm 0.10$ \\
    & $K^{*0} \eta'$
        & $7.8^{+7.7}_{-5.7} \; (<24)$ \cite{Richichi:1999kj}
        & $3.2^{+1.8}_{-1.6} \pm 0.9 \; (<6.4)$ \cite{Aubert:2003ru}
        & $<20$ \cite{Aihara2003}
        & $<6.4$ \\
    & $\rho^- K^+$
        & $16.0^{+7.6}_{-6.4} \pm 2.8 \; (<32)$ \cite{Jessop:2000bv}
        & $7.3^{+1.3}_{-1.2} \pm 1.3$ \cite{Aubert:2003wr}
        & $15.1^{+3.4+1.4+2.0}_{-3.3-1.5-2.1}$ \cite{Belle0317}
        & $9.0 \pm 2.3~(S=1.41)$ \\
    &   & -
        & $0.18 \pm 0.12 \pm 0.08$ \cite{Jawahery:2003}
        & $0.22^{+0.22+0.06}_{-0.23-0.02}$ \cite{Belle0317}
        & $0.19 \pm 0.12$ \\
    & $\rho^0 K^0$
        & $<39$ \cite{Asner:1996hc}
        & -
        & $<12.4$ \cite{Huang:2002ev}
        & $<12.4$ \\
    & $\omega K^0$
        & $10.0^{+5.4}_{-4.2}\pm1.4 \; (<21)$ \cite{Jessop:2000bv}
        & $5.3^{+1.4}_{-1.2}\pm0.5$ \cite{Aubert:2003fa}
        & $4.0^{+1.9}_{-1.6} \pm 0.5$ \cite{Abe2003}
        & $5.2 \pm 1.1$ \\
    & $\phi K^0$
        & $5.4^{+3.7}_{-2.7} \pm 0.7 \; (<12.3)$ \cite{Briere:2001ue}
        & $7.6^{+1.3}_{-1.2} \pm 0.5$ \cite{Aubert:2003tk}
        & $9.0^{+2.2}_{-1.8} \pm 0.7$ \cite{Chen:2003}
        & $7.8 \pm 1.1$ \\
\end{tabular}
\end{ruledtabular}
\leftline{$^a$Value utilized in order to stabilize fits.  See text.}
\end{table*}

%
\begin{table*}
\caption{Summary of predicted contributions to $\Delta S = 0$ decays of $B$
  mesons to one vector and one pseudoscalar mesons.  Amplitude magnitudes
  $|A_{\rm exp}|$ extracted from experiments are quoted in units of eV.
  The results are based on a fit with $p'_V = - p'_P$
and $\gamma\simeq65^{\circ}$ (see Table V).
\label{tab:dS0}}
\begin{ruledtabular}
\begin{tabular}{llcccccc}
 & Mode & Amplitudes & $|\cal{T}|$$^a$ & $|\cal{P}|$$^a$ 
 & $p_c$ (GeV) & $|A_{\rm exp}|^b$ & $A_{CP}$ \\ 
\hline
$B^+ \to$
    & ${\lbar K}^{*0} K^+$
        & $p_P$
        & 0 & 7.5 & 2.539 & $<24.1$ & - \\
    & $K^{*+} \bar K^0$
        & $p_V$
        & 0 & 7.5  & 2.539 & - & - \\
    & $\rho^0 \pi^+$ 
        & $-\frac{1}{\sqrt{2}}(t_V+c_P+p_V-p_P)$
        & 25.1 & 10.6 & 2.582 & $31.4\pm1.9$ & $-0.17 \pm 0.11$ \\
    & $\rho^+ \pi^0$ & $-\frac{1}{\sqrt{2}}(t_P+c_V+p_P-p_V)$
        & 39.7 & 14.4 & 2.582 & $34.5 \pm 4.2$ & $0.23 \pm 0.17$ \\
    & $\rho^+ \eta$ 
        & $-\frac{1}{\sqrt{3}}(t_P+c_V+p_P+p_V+s_V)$
        & 32.4 & 2.1$^c$ & 2.554 & $31.2 \pm 4.7$ & $0.06 \pm 0.29$ \\
    & $\rho^+ \eta'$ 
        & $\frac{1}{\sqrt{6}}(t_P+c_V+p_P+p_V+4s_V)$
        & 22.9 & 0.7$^c$ & 2.493 & $38.7 \pm 6.6$ & - \\
    & $\omega \pi^+$ 
        & $\frac{1}{\sqrt{2}}(t_V+c_P+p_P+p_V+2s_P)$
        & 25.1 & 0.006$^d$ & 2.580 & $25.3 \pm 2.3$ & $0.11 \pm 0.21$ \\
    & $\phi \pi^+$
        & $s_P$
        & 0 & 0.009$^d$ & 2.539 & $<6.7$ & - \\
\hline
$B^0 \to$
    & $\ol K^{*0} K^0$
        & $p_P$ 
        & 0 & 7.5 & 2.539 & - & - \\
    & $K^{*0} \ol K^0$
        & $p_V$
        & 0 & 7.5 & 2.539 & - & - \\
     & $\rho^- \pi^+$
        & $-(t_V+p_V)$
        & 30.3 & 7.5 & 2.582 & $34.4 \pm 3.4^{~e}$ 
        & $-0.54 \pm 0.19$ \\
     & $\rho^+ \pi^-$
        & $-(t_P+p_P)$
        & 43.1 & 7.5 & 2.582 & $40.1 \pm 3.2^{~e}$
        & $-0.16 \pm 0.15$ \\
    & $\rho^0 \pi^0$ 
        & $\frac12(-c_P-c_V+p_P+p_V)$
        & 9.2$^f$ & 2.7 & 2.582 & $<17.1$ & - \\
    & $\rho^0 \eta$ 
        & $\frac{1}{\sqrt{6}}(c_P-c_V-p_P-p_V-s_V)$
        & 3.2$^f$ & 1.5$^c$ & 2.554 & $<25.4$ & - \\
    & $\rho^0 \eta'$ 
        & $\frac{1}{2\sqrt{3}}(c_V-c_P+p_P+p_V+4s_V)$
        & 2.3$^f$ & 0.5$^c$ & 2.493 & $<38.0$ & - \\
    & $\omega \pi^0$
        & $\frac{1}{2}(c_P-c_V+p_P+p_V+2s_P)$
        & 4.0$^f$ & 2.7$^d$ & 2.580 & $<14.9$ & - \\
    & $\omega \eta$
        & $-\frac{1}{\sqrt{6}}(c_P+c_V+p_P+p_V+2s_P+s_V)$
        & 7.5$^f$ & 1.5$^{c,d}$ & 2.552 & $< 37.6$ & - \\
    & $\omega \eta'$
        & $\frac{1}{2\sqrt{3}}(c_P+c_V+p_P+p_V+2s_P+4s_V)$
       & 5.3$^f$ & 0.5$^{c,d}$ & 2.491 & $ < 85.1$ & - \\
    & $\phi \pi^0$
        & $\frac{1}{\sqrt{2}}s_P$
        & 0 & 0.006$^d$ & 2.539 & $<24.3$ & - \\
    & $\phi \eta$
        & $-\frac{1}{\sqrt{3}}s_P$
        & 0 & 0.005$^d$ & 2.511 & $ < 32.8$ & - \\
    & $\phi \eta'$
        & $\frac{1}{\sqrt{6}}s_P$
        & 0 & 0.004$^d$ & 2.447 & $ < 11.1$ & - \\
\end{tabular}
\end{ruledtabular}
\leftline{$^a$ $\cal{T}$ is the sum of all tree and color-suppressed
amplitudes that contribute to a process.}
\leftline{$\ \ \, \cal{P}$ is the sum of all penguin amplitudes, including
electroweak ones.}
\leftline{$^b$ $|A_{\rm exp}|$ is defined by Eq.~(\ref{eq:width}) as an 
amplitude related to a $CP$-averaged branching ratio quoted in Table I.}
\leftline{$^c$ No $S_V$ contribution included.  $^d$ No $S_P$ contribution
included.} 
\leftline{$^e$ Based on $CP$-averaged branching ratios quoted in Table I.}
\leftline{$^f$ Takes account of the relative phase between $C_P$ and $C_V$ 
amplitudes.}
\end{table*}
%

\begin{table*}
\caption{Same as Table \ref{tab:dS0} for $|\Delta S| = 1$ decays of $B$
  mesons.
\label{tab:dS1}}
\begin{ruledtabular}
\begin{tabular}{llcccccc}
 & Mode & Amplitudes & $|\cal{T}'|$ & $|\cal{P}'|$
 & $p_c$ (GeV) & $|A_{\rm exp}|$ & $A_{CP}$ \\ 
\hline
$B^+ \to$
    & $K^{*0} \pi^+$
        & $p'_P$
        & 0 & 32.6 & 2.561 & $31.2 \pm 2.4$ & - \\
    & $K^{*+} \pi^0$ 
        & $-\frac{1}{\sqrt{2}}(t'_P+c'_V+p'_P)$
        & 9.4 & 39.7 & 2.562 & $<58.1$ & - \\
    & $K^{*+} \eta$ 
        & $-\frac{1}{\sqrt{3}}(t'_P+c'_V+p'_P-p'_V+s'_V)$
        & 7.7 & 46.7$^a$ & 2.534 & $53.4 \pm 3.5$ & $0.10 \pm 0.12$ \\
    & $K^{*+} \eta'$ 
        & $\frac{1}{\sqrt{6}}(t'_P+c'_V+p'_P+2p'_V+4s'_V)$
        & 5.4 & 16.5$^a$ & 2.472 & $<62.8$ & - \\
    & $\rho^0 K^+$
        & $-\frac{1}{\sqrt{2}}(t'_V+c'_P+p'_V)$
        & 6.9 & 22.9 & 2.559 & $21.2\pm2.1$ & - \\
    & $\rho^+ K^0$
        & $p'_V$
        & 0 & 32.6 & 2.559 & $<72.3$ & - \\
    & $\omega K^+$ 
        & $\frac{1}{\sqrt{2}}(t'_V+c'_P+p'_V+2s'_P)$ 
        & 6.9 & 23.0$^b$ & 2.557 & $24.3 \pm 1.8$ & $-0.003 \pm 0.122$ \\
    & $\phi K^+$
        & $p'_P+s'_P$ 
        & 0 & 32.5$^b$ & 2.516 & $31.6 \pm 1.6$ & $0.030 \pm 0.072$ \\
\hline
$B^0 \to$
    & $K^{*+} \pi^-$
        & $-(t'_P+p'_P)$ 
        & 10.3 & 32.6 & 2.562 & $42.4 \pm 5.2$ & $0.26 \pm 0.35$ \\
    & $K^{*0} \pi^0$
        & $-\frac{1}{\sqrt{2}}(c'_V - p'_P)$
        & 2.1 & 6.3 & 2.562 & $<20.3$ & - \\
    & $K^{*0} \eta$
        & $-\frac{1}{\sqrt{3}}(c'_V + p'_P-p'_V + s'_V)$
        & 1.7 & 46.7$^a$ & 2.534 & $46.0 \pm 2.6$ & $0.05 \pm 0.10$ \\
    & $K^{*0} \eta'$
        & $\frac{1}{\sqrt{6}}(c'_V + p'_P + 2p'_V + 4s'_V)$
        & 1.2 & 16.5$^a$ & 2.471 & $<39.8$ & - \\
    & $\rho^- K^+$
        & $-(t'_V+p'_V)$
        & 8.5 & 32.6 & 2.560 & $32.4 \pm 4.2$ & $0.19 \pm 0.12$ \\
    & $\rho^0 K^0$
        & $\frac{1}{\sqrt{2}}(p'_V-c'_P)$
        & 0.9 & 23.1 & 2.559 & $<38.2$ & - \\
    & $\omega K^0$
        & $\frac{1}{\sqrt{2}}(c'_P + p'_V + 2s'_P)$
        & 0.9 & 23.0$^b$ & 2.557 & $24.6 \pm 2.6$ & - \\
    & $\phi K^0$
        & $p'_P+s'_P$ 
        & 0 & 32.5$^b$ & 2.516 & $30.5 \pm 2.1$ & - \\
\end{tabular}
\end{ruledtabular}
\leftline{$^a$ No $S'_V$ contribution included.  $^b$ No $S'_P$ contribution
included.}
\end{table*}

\section{EXTRACTING AMPLITUDES \label{sec:extract}}

In the present section we show how a global fit to decay rates and $CP$
asymmetries can determine many (though not all) of the invariant amplitudes
governing $B \to VP$ decays.  We shall be able to determine from experimental
data their magnitudes and relative strong phases and the weak phase $\gamma$.
We shall assume a universal ratio $p'_V/p'_P = -ce^{i \phi}$, initially
assuming $c=1$ and $\phi=0$ in accord with Ref.\ \cite{HJLP}, presenting also
results with arbitrary $c$ and $\phi$.  We interpret
$\phi$ as a relative {\it strong} phase between $p'_V$ and $-p'_P$, so that
it does not change sign under $CP$-conjugation.  We now explain in some
detail the inputs and fit parameters.

We base the present fit on the following processes
(see Tables \ref{tab:dS0} and \ref{tab:dS1}):

\begin{itemize}

\item The $B^+ \to K^{*0} \pi^+$ amplitude involves $|p'_P|$ alone.  The
decay rate provides one data point.
No $CP$ asymmetry is expected or seen.

\item The decays $B^0 \to \rho^- \pi^+$ and $\ob \to \rho^+ \pi^-$
(equivalently, their $CP$-averaged branching ratio and $CP$ asymmetry quoted in
Table \ref{tab:dS0data}) involve $t_V$ and $p_V$.  These processes thus provide
two data points.

\item The decays $B^0 \to \rho^+ \pi^-$ and $\ob \to \rho^- \pi^+$ involve
$t_P$ and $p_P$ and provide two data points.

\item In the time-dependent study of $(B^0,\ob) \to \rho^\mp \pi^\pm$,
the asymmetry parameters $S_{+-}$ and $S_{-+}$, to be defined at the end of
this section, provide two more data points.  (Other time-dependent parameters
are related to those already included.)
 
\item The decays $B^0 \to \rho^- K^+$ and $\ob \to \rho^+ K^-$ involve $t'_V$
and $p'_V$ and provide two data points, since the $CP$-averaged decay rate
and $CP$ asymmetry have been presented.

\item The decays $B^0 \to K^{*+} \pi^-$ and $\ob \to K^{*-} \pi^+$ involve
$t'_P$ and $p'_P$ and similarly provide two data points.

\item The decays $B \to K^* \eta$ (for both charge states) play an important
role in constraining the phase $\phi$ of $-p'_V/p'_P$, since this phase must
be small in order that $p'_P$ and $p'_V$ contribute constructively to the
large decay rate, as anticipated in Ref.\ \cite{HJLP}.  We include two
decay rates and two $CP$ asymmetries, adding a total of four data points.
Since our scheme predicts a very small $CP$ asymmetry for $B^0 \to K^{*0}
\eta$, the parameters of the fit will not be affected by this observable.

\item The rate and $CP$ asymmetry for $B^+ \to \rho^0 \pi^+$ and
$B^+ \to \omega \pi^+$ have been measured.  The two decay rates are
dominated by $t_V$ but provide some information about the magnitude of the
amplitude $c_P$, about which we shall have more to say below.  These processes
thus add four more data points to our fits.

\item The rate and $CP$ asymmetry for $B^+ \to \rho^+ \pi^0$ have been
measured, adding two data points.

\item The decay rates for $B \to \phi K$ (both charge states) have been
measured.
The corresponding decay widths are expected to be equal. They are measured 
to be within 7\% when one takes into account the difference in lifetimes of 
the $B^+$ and $B^0$ mesons.
[Note added:  The branching ratio $\cB(B^0 \to \phi K^0)$ quoted in Table
\ref{tab:dS1data} has been updated \cite{Aubert:2003tk}.  The central values
of the world-averaged decay widths now are exactly equal.]
We include 
the $B \to \phi K$ decay rates as two
more data points.  Since both the amplitudes $p'_P$ and $s'_P$ contributing to
these processes are expected to have the same weak phase, we predict zero $CP$
asymmetry in any $B \to \phi K$ decay.  This is certainly true for the charged
mode, whose $CP$ asymmetry we include as another data point.

\item Taking the average of BaBar and Belle values \cite{Browder:2003}, we find
the time-dependent parameters in $B^0 \to \phi K_S$ to be $S_{\phi K_S} =
-0.147 \pm 0.697~(S=2.11)$ and $A_{\phi K_S} = 0.046 \pm 0.256~(S=1.08)$,
whereas we predict the standard model values $(\sin 2 \beta, 0)$.  The average
of BaBar and Belle determinations via the subprocess $\bar b \to \bar c c \bar
s$ is $\sin 2 \beta = 0.736 \pm 0.049$ \cite{Browder:2003}.  The parameter
$A_{\phi K_S}$ is equivalent to the direct $CP$ asymmetry $A_{CP}(B^0 \to \phi
K_S)$.  (The corresponding asymmetry for $B^+ \to \phi K^+$ is seen in Table II
to be very small.) These observables thus contribute 
$\Delta \chi^2 = (1.61, 0.03)$
to our fit, without affecting the fit parameters.  We include them
in our $\chi^2$ total, adding two more data points.  In view of the large
$S$-factor, contributing to a considerable amplification of the experimental
error in $S_{\phi K_S}$, it is premature to regard the deviation of this
observable from its standard model expectation as a signal of new physics.
In Ref.\ \cite{Chiang:2003jn} we discussed some scenarios which could give rise
to such a deviation.  One should add a penguin amplitude (e.g., for $\bar b \to
\bar s s \bar s$) with arbitrary magnitude and weak and strong phases to the
present global fit to see if one can describe all the $B \to VP$ data with any
greater success.

\item Both a decay rate and a $CP$ asymmetry have been presented for
$B^+ \to \omega K^+$, while we are aware only of a decay rate for $B^0 \to
\omega K^0$.  We thus add three more data points for these processes.

\item The BaBar Collaboration has recently reported observation of the decay
modes $B^+ \to \rho^+ \eta$ and $B^+ \to \rho^+ \eta'$ \cite{Aubert:2003ru} at
levels indicating a significant role for the $C_V$ amplitude.  We include
the branching ratios for these processes as averages between the BaBar
and older CLEO \cite{Richichi:1999kj} values.  In addition we include the
new BaBar value of 
$A_{CP}(B^+ \to \rho^+ \eta)$.

\item The decay rate for $B^+ \to \rho^0 K^+$ was recently measured by the 
Belle collaboration with high significance \cite{Belle0338}. We include 
the average between the Belle result and the previous measurements by BaBar 
and CLEO as another data point.

\item Although only an upper limit exists so far for $\cB(B^0 \to K^{*0}
\pi^0)$, we use the Belle central value and error 
\cite{Belle0317}
in order to enforce this upper limit in the fits.

\end{itemize}

The grand total of fitted data points is thus 34, including some quantities
such as $A_{CP}(B^+ \to \phi K^+)$, $S_{\phi K_S}$ and $A_{\phi K_S}$ which do
not affect our fit.  We now count the parameters of the fit.

\begin{itemize}

\item The amplitude $p'_P$ is taken to have a strong phase of $\pi$ by
definition.  Its weak phase, since it is dominated by $-V^*_{cb} V_{cs}$
(see the discussion at the end of Sec.\ \ref{sec:not}), also is $\pi$,
so we will have $p'_P$ real and positive.  This choice will be seen in
our favored solution to entail tree amplitudes with positive real and
small imaginary parts (when their weak phases are neglected),
in accord with expectations from factorization
\cite{Beneke:2003}.  The corresponding strangeness-preserving amplitude
$p_P$ is determined by $p_P = (V_{cd}/V_{cs}) p'_P = - \bl p'_P$, where
$\bl \equiv \lambda/(1 - \frac{\lambda^2}{2}) = 0.230$ and $\lambda = 0.224$
\cite{Battaglia:2003in}.  Thus the weak phase of $p_P$ in the present
convention is zero, while its strong phase is $\pi$.  When we assume that
$p'_V/p'_P = p_V/p_P = -1$, as suggested in Ref.\ \cite{HJLP} and as done
in Refs. \cite{VPUP} and \cite{Chiang:2001ir}, we will have one free
parameter $|p'_P|$.  More generally, we shall consider fits with this
ratio real or complex, adding one or two new parameters $c$ and
$\phi$ defined by $p'_V/p'_P = p_V/p_P = -c$ or $p'_V/p'_P = p_V/p_P =
-c e^{i \phi}$.  We do not introduce SU(3) breaking in penguin amplitudes.

\item 
The magnitudes $|t_{P,V}|$ of the tree amplitudes and 
strong relative phases $\delta_{P,V}$ between them and the corresponding
penguin amplitudes $p_{P,V}$ are free parameters:  four in all.  We relate
strangeness-changing tree amplitudes to those for $\Delta S = 0$ using
ratios of decay constants:  $t'_V = \bl (f_K/f_\pi) t_V \simeq
0.281 t_V$ and $t'_P = \bl (f_{K^*}/f_\rho) t_P \simeq 0.240 t_P$.
We thus assume the same relative strong phases between tree and penguin
amplitudes in $\Delta S = 0$ and $|\Delta S| = 1$ processes.
We use the following values of the decay constants \cite{Chiang:2001ir,PDG}:
$f_\pi = 130.7$~MeV, $f_K = 159.8$~MeV, $f_\rho = 208$~MeV, 
$f_{K^*} = 217$~MeV.

\item The weak phase $\gamma$ of $t_V$ and $t_P$ is a free parameter.
We assume it to be the same for both tree amplitudes.

\item 
We take the electroweak penguin amplitude $P'_{EW,P}$ to have the
same strong and weak phases as $p'_P$. Then the electroweak penguin 
contribution to $s'_P$, $-\frac13 P'_{EW,P}$, interferes destructively
with $p'_P$, as was anticipated by explicit calculations \cite{EWPp}. 
Thus $-s'_P/p'_P$ is one real positive parameter.  We ignore any contribution
from the singlet penguin $S'_P$, which we expect to involve gluonic coupling to
SU(3) flavor-singlet components of vector mesons and thus to be suppressed
by the Okubo-Zweig-Iizuka (OZI) rule.  We did not find a stable fit if we
allowed the strong phase of the EWP contribution to $s'_P$ to vary.
The contribution $-\frac{1}{3}P'_{EW,P}$ appearing in $s'_P$ then implies a
corresponding contribution $+P'_{EW,P}$ in $c'_P$.

\item The term $c_P$ appears to play a key role in accounting for the
deviation of the $B^+ \to \rho^0 \pi^+$ and $B^+ \to \omega \pi^+$ decay
rates from the predictions based on $t_V$ alone.  It contains two terms:
a term $C_P$ which we choose to have the same weak and strong phases
as $t_V$, and a small EWP contribution $P_{EW,P}= - \bl P'_{EW,P}$ associated
with the term taken in $s'_P$ above. Thus $C_P/t_V$ will be one additional real
parameter, which will turn out to be positive in all our fits. 

\item We include a contribution from the amplitude $C_V$, motivated by the
large $B^+ \to \rho^+ \eta$ and $B^+ \to \rho^+ \eta'$ branching ratios.  
We choose $C_V$ to have the same strong and weak phases as $t_P$.
We do not introduce SU(3) breaking in $C_P$ and $C_V$ amplitudes and assume
$C'_P= \bl C_P$, $C'_V= \bl C_V$.

\item We take the electroweak penguin contribution to $s'_V$, $-\frac{1}{3}
P'_{EW,V}$, to have the same strong and weak phases as $p'_V$.  This
contribution then implies corresponding contributions $+P'_{EW,V}$ in $c'_V$
and $+P_{EW,V}=-\bl P'_{EW,V}$ in $c_V$.
The apparent suppression of the decay $B^0 \to K^{*0} \pi^0$ (see Sec.\ VI)
suggests that $P'_{EW,V}$ and $p'_P$ are interfering destructively in this
process, implying constructive interference in both charge states of $B \to
K^* \eta$.
We ignore any contribution from the singlet penguin $S'_V$ in the
absence of information about its magnitude and phase.

\end{itemize} 

There are thus ten, eleven, or twelve parameters to fit 34 data points,
depending on the assumption for the $p'_V/p'_P$ ratio: $-1$, real, or complex.
In fact, not all partial decay rates are independent, as we must have the
following equalities between rate differences \cite{Deshpande:2000jp}:
\bea
\Gamma(B^0 \to \rho^- \pi^+)  & - & \Gamma(\ob \to \rho^+ \pi^-) = \nonumber \\
 (f_\pi/f_K) [\Gamma(\ob \to \rho^+ K^-)   & - & \Gamma(B^0 \to \rho^- K^+)]~,
 \label{eqn:rateV} \\
\Gamma(B^0 \to \rho^+ \pi^-)  & - & \Gamma(\ob \to \rho^- \pi^+) = \nonumber \\
 (f_\rho/f_{K^*})[ \Gamma(\ob \to K^{*-} \pi^+) & - &
 \Gamma(B^0 \to K^{*+} \pi^-)]~.
 \label{eqn:rateP}
\eea
When transcribed into relations among branching ratios, these read,
respectively,

\be
 (11.1 \pm 3.8) \times 10^{-6} \stackrel{?}{=} (2.8 \pm 1.9)
 \times 10^{-6}~,
 \label{eqn:brV}
\ee
\be
 (4.4 \pm 4.1) \times 10^{-6} \stackrel{?}{=} (7.6 \pm 10.4)
 \times 10^{-6}~.
 \label{eqn:brP}
\ee
The first of these is violated at the $2 \sigma$ level.  One reaps little gain
in relaxing the assumption $p'_V = - p'_P$ since these relations are expected
to hold regardless of $p'_V/p'_P$.

When we assume $p'_V = - p'_P$, the specific expressions entering our fits
include
\bea
A(B^+ \to K^{*0} \pi^+)  & = & |p'_P| \\
A(B^0 \to \rho^- \pi^+) & = & -\bl |p'_P| - |t_V| e^{i(\delta_V + \gamma)} \\ 
A(B^0 \to \rho^- K^+)   & = & |p'_P| 
 - \bl \frac{f_K}{f_\pi} |t_V| e^{i(\delta_V + \gamma)} \\
A(B^0 \to \rho^+ \pi^-) & = & \bl |p'_P| + |t_P| e^{i(\delta_P + \gamma)}\\
A(B^0 \to K^{*+} \pi^-) & = & - |p'_P|
 + \bl \frac{f_{K^*}}{f_\rho} |t_P| e^{i(\delta_P + \gamma)}~.
\eea
The phase convention is such that $\delta_{V,P} = 0$ corresponds to $t_{P,V}$
having a phase of $\gamma$ with respect to $p_{P,V}$.  Amplitudes associated
with the charge-conjugate modes can be obtained by flipping the sign of
$\gamma$ in the above expressions.  The expressions lead to the rate relations
(\ref{eqn:rateV}) and (\ref{eqn:rateP}) if one squares them and takes
appropriate differences.

We now discuss two additional parameters in $(B^0,\ob) \to \rho^\pm \pi^\mp$
which provide further constraints.  These are measured in a time-dependent
study by the BaBar Collaboration \cite{Jawahery:2003}:

\bea
S_{\rho \pi} & = & -0.13 \pm 0.18 \pm 0.04~~,\\
\Delta S_{\rho \pi} & = & 0.33 \pm 0.18 \pm 0.03~~,
\eea
which are related to the parameters $S_{+-}$ and $S_{-+}$ by
\bea
S_{\rho \pi} & = & (S_{+-} + S_{-+})/2~~, \\
\Delta S_{\rho \pi} & = & (S_{+-} - S_{-+})/2~~,
\eea
where
\bea
S_{+-} & \equiv & \frac{2 {\rm Im} \lambda^{+-}}{1 + |\lambda^{+-}|^2}~~,\\
S_{-+} & \equiv & \frac{2 {\rm Im} \lambda^{-+}}{1 + |\lambda^{-+}|^2}~~,\\
\eea
\bea
\lambda^{+-} & \equiv & \frac{q}{p}
\frac{A(\bar B^0 \to \rho^+ \pi^-)}{A(B^0 \to \rho^+ \pi^-)}~~,\\
\lambda^{-+} & \equiv & \frac{q}{p}
\frac{A(\bar B^0 \to \rho^- \pi^+)}{A(B^0 \to \rho^- \pi^+)}~~
\eea
and $q/p = e^{-2 i \beta}$ with $\beta = 23.7^\circ$.
Since our fits predict the phases and magnitudes of all the relevant decay
amplitudes, it is a simple matter to calculate the $S$'s.  They provide
crucial information on the relative strong phases of $t_V$ and $t_P$, among
other things.  As mentioned, other observables $C_{\rho \pi}$, $\Delta
C_{\rho \pi}$, and $A_{\rho \pi}$ as defined in Refs.\ \cite{Hocker} and
\cite{Aubert:2003wr} are related to information we already use in our fits and
need not be considered separately.  (See the Appendix.)

To see explicitly the constraints provided by $S_{\rho \pi}$ and
$\Delta S_{\rho \pi}$ it is helpful to calculate them in the limit in which
the small penguin contributions to the $B \to \rho \pi$ amplitudes can be
neglected.  Defining $r \equiv |t_V/t_P|$ and $\delta \equiv {\rm Arg}
(t_V/t_P)$, one finds $S_{\pm \mp} = 2r \sin(2 \alpha \pm \delta)/(1+r^2)$
and
\be
S_{\rho \pi} = \frac{2r}{1+r^2} \sin 2 \alpha \cos \delta,~
\Delta S_{\rho \pi} = \frac{2r}{1+r^2} \cos 2 \alpha \sin \delta~.
\ee
Both these quantities are small experimentally, consistent with
solutions in which $\alpha = \pi - \beta - \gamma$ is near $\pi/2$ while
$\delta$ is near zero or $\pi$.
The non-zero central value of $\Delta S_{\rho \pi}$, in our favored solution
to be discussed below, combines with a value of $\cos 2 \alpha$ near $-1$
and other constraints to favor a small negative value of $\delta$.

\section{TESTS OF THE $p'_V = -p'_P$ ASSUMPTION \label{sec:test}}

In this section, we note that the relation $p'_V = -p'_P$, proposed in Ref.\
\cite{HJLP} and used in previous discussions \cite{VPUP,Chiang:2001ir}, can
be tested experimentally, and discuss the status of such tests.

As described before, information on $|p'_P|$ is directly obtained from the $B^+
\to K^{*0} \pi^+$ decay rate, which is predicted to be equal to that of its
$CP$ conjugate.  (The absence of a $CP$ asymmetry in this mode is one test of
the present picture.)  In principle, one can also extract $|p'_V|$ directly
from the $B^+ \to \rho^+ K^0$ rate (for which one also expects a vanishing $CP$
asymmetry).  Currently, only the CLEO group has reported an upper bound on the
averaged branching ratio, $\lbar{\cal B}(\rho^+ K^0)< 48 \times 10^{-6}$ 
based on a sample of $2.6$~million $B\bar{B}$ pairs.
The present BaBar and Belle sample of approximately 100 times as much data
would enable this mode to be observed at the predicted branching ratio of
$\lbar{\cal B}(\rho^+ K^0) \simeq 12.6 \times 10^{-6}$ with good significance.

An indirect way to test $p'_P = -p'_V$ is to compare the $\omega K^0$ and $\phi
K^0$ modes.  Using $c'_P = C'_P + P'_{EW,P}$ and $s'_P = S'_P - P'_{EW,P}/3$
[Eq.~(\ref{eq:dict})] and the facts that $C'_P$ is smaller than $P'_{EW,P}$ by
a factor of about $0.22$ (see Table I in Ref.~\cite{Chiang:2001ir}) and that
$S'_P$ is OZI-suppressed, we can safely neglect them and have
\bea
A(\omega K^0)
&=& \frac{1}{\sqrt{2}} \left( p'_V + \frac13 P'_{EW,P} \right) ~, \\
A(\phi K^0)
&=& p'_P - \frac13 P'_{EW,P} ~.
\eea
Therefore, the amplitude magnitudes are related by a factor of $\sqrt{2}$
if $p'_P = -p'_V$.  Current data, using the average for $B^+ \to \phi
K^+$ and $B^0 \to \phi K^0$ amplitudes (expected to be equal within our
approximations) show that 
$\sqrt{2}|A(B^0 \to \omega K^0)/A(B \to \phi K)| = 1.13 \pm 0.13$, consistent
with 1.
[Note added: The updated branching ratio $\cB(B^0 \to \phi K^0)$
\cite{Aubert:2003tk} results in a slightly smaller ratio of amplitudes: 
$\sqrt{2}|A(B^0 \to \omega K^0)/A(B \to \phi K)| = 1.11 \pm 0.13$.]

Observation of the $\rho^{\pm} K^0$ mode and more precise determination of
$\omega K^0$ and $\phi K^0$ modes thus will be very helpful in justifying the
assumption of $p'_P = -p'_V$.  However, we find in the next section that global
fits to data in which this assumption is relaxed are not very different from
those in which $p'_P = -p'_V$ is assumed.

\section{DISCUSSION OF PREDICTIONS
\label{sec:var}}

\begin{figure}[h]
\includegraphics[width=.5\textwidth]{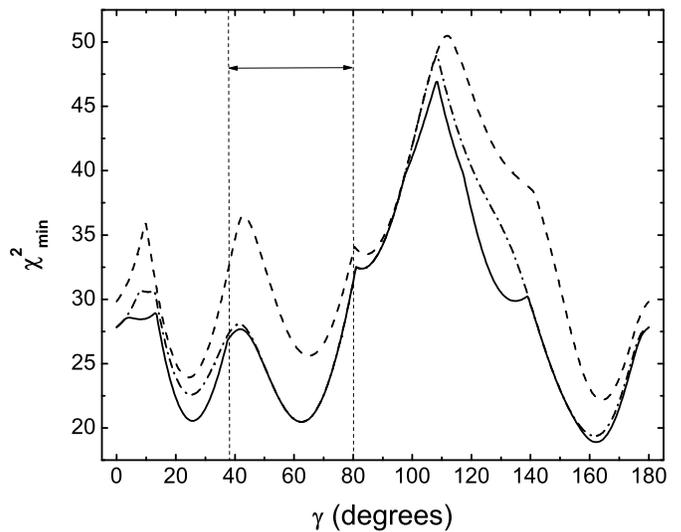}
\caption{$(\chi^2)_{\rm min}$, obtained by minimizing over all remaining fit
parameters, as a function of the weak phase $\gamma$.  Dash-dotted curve:
$p'_V/p'_P = -1$ (24 d.o.f.); dashed curve: $p'_V/p'_P$ real (23 d.o.f.);
solid curve: $p'_V/p'_P$ complex (22 d.o.f.).
\label{fig:gc}}
\end{figure}

Plots of $\chi^2$ as a function of $\gamma$ for three fits are shown in
Fig.\ \ref{fig:gc}.  Three local minima are found, around $\gamma = 26^\circ$,
$63^\circ$, and $162^\circ$.  The fit with $p'_V/p'_P$ real gives a
$\chi^2$ very similar to that with $p'_V/p'_P = -1$ for $\gamma \simeq
26^\circ$ and to that with $p'_V/p'_P$ complex for the other two minima, so
we shall not consider it further.  The magnitudes of individual amplitudes and
the strong phases determined in the fits with $p'_V/p'_P = -1$ and $p'_V/p'_P$
complex are compared with one another for the three local minima in Table
\ref{tab:amps}.  The corresponding predictions of these fits are compared with
one another and with experiment in Tables \ref{tab:pred0} and \ref{tab:pred1}.
\null

\null

In the absence of information on $S_{\rho \pi}$ and $\Delta S_{\rho \pi}$,
Fig.\ \ref{fig:gc} would be symmetric under $\gamma \to \pi - \gamma$, since
all other observables would be unchanged under a simultaneous change of strong
phases $\delta_{P,V} \to \pi - \delta_{P,V}$.  The time-dependent $CP$
asymmetries in $B \to \rho^\mp \pi^\pm$ break this symmetry.  Nonetheless,
considerable ambiguity remains, and it is necessary to appeal both to theory
and to other experiments to resolve it.

\begin{figure}
\includegraphics[width=.5\textwidth]{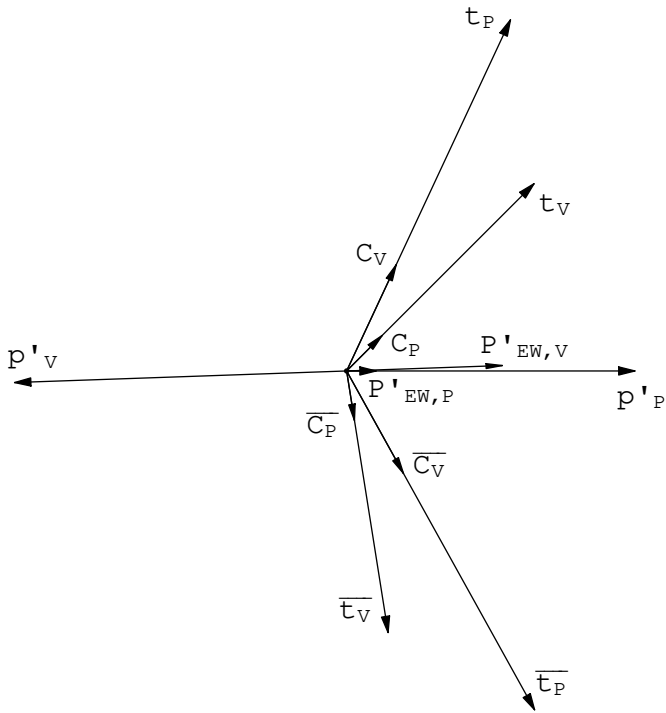}
\caption{Magnitudes and phases of dominant invariant amplitudes in solution
with $\gamma \simeq 63^\circ$ and complex $p'_V/p'_P$.  Other amplitudes are
given by $p_{(P,V)} = -0.230\, p'_{(P,V)}$, 
$P_{EW,(P,V)}=-0.230\, P'_{EW,(P,V)}$,
$t'_P = 0.240\, t_P$, $t'_V =0.281\, t_V$, $C'_{(P,V)}=0.230\, C_{(P,V)}$.
\label{fig:phas}}
\end{figure}

We shall concentrate upon solutions consistent with $\gamma$ in the range
allowed by fits \cite{CKMf} to other observables, $38^\circ < \gamma < 80
^\circ$ at the 95\% confidence level.  Solutions outside this range not
only conflict with these fits, but have large final-state phases (mod $\pi$)
which are unlikely in QCD factorization \cite{Beneke:2003}.  
We shall point
out ways in which a distinction among solutions can be made experimentally. 
Magnitudes and phases of the dominant invariant amplitudes in the
solution with $\gamma \simeq 63^\circ$ and complex $p'_V/p'_P$ are shown
in Fig.\ \ref{fig:phas}.  Using this figure and Tables \ref{tab:dS0} and
\ref{tab:dS1} one can see whether a given process involves constructive
or destructive tree-penguin interference.

Errors for all solutions with complex $p'_V/p'_P$ have been estimated
using a Monte Carlo method in which parameter sets were generated leading to
$\chi^2$ values no more than 1 unit above the minimum, and the spread in
predictions was studied.  For any prediction depending on a single parameter,
the error in that parameter was used to obtain the error in the prediction.

\begin{table*}
\caption{Comparison of parameters extracted in fits to branching ratios and
$CP$ asymmetries under various assumptions.  Values of the topological
amplitudes are quoted in units of eV.  Probabilities are those for $\chi^2$ to
exceed the value shown for the indicated number of degrees of freedom.
\label{tab:amps}}
\begin{ruledtabular}
\begin{tabular}{lcccccc}
         & \multicolumn{6}{c}{Value in fit} \\
Quantity & $| \longleftarrow$ & $p'_V/p'_P = -1$ & $\longrightarrow |$
     & $| \longleftarrow$ & $p'_V/p'_P = -ce^{i \phi}$ & $\longrightarrow |$ \\
\hline
$\gamma$ & $(24\pm5)^\circ$ & $(65\pm6)^\circ$ & $(164\pm5)^\circ$
      & $(26\pm5)^\circ$ & $(63\pm6)^\circ$ & $(162^{+5}_{-6})^\circ$ \\ 
\hline
$\phi$   & 0 (input) & 0 (input) & 0 (input)
   & $(35^{+12}_{-21})^\circ$ & $(2\pm18)^\circ$ & $(-20^{+31}_{-23})^\circ$ \\
$|p'_P|$ & $32.4^{+1.4}_{-1.5}$ & $32.6\pm1.5$ & $32.4^{+1.4}_{-1.5}$
        & $32.5^{+1.7}_{-1.6}$ & $32.2^{+1.4}_{-1.5}$ & $32.0\pm1.6$ 
\\
$|p'_V|$ & $32.4^{+1.4}_{-1.5}$ & $32.6\pm1.5$ & $32.4^{+1.4}_{-1.5}$
        & $31.3^{+3.6}_{-3.0}$ & $37.1^{+2.3}_{-2.4}$ & $35.3^{+3.4}_{-3.6}$ \\
$|s'_P|^{~a}$ & $0.9\pm1.3$ & $0.0\pm1.3$ & $0.5\pm1.3$
              & $1.7^{+1.8}_{-1.6}$ & $1.1\pm1.3$ & $1.1^{+1.7}_{-1.5}$ \\
$|s'_V|^{~a}$ & $6.2^{+2.2}_{-2.1}$ & $7.9^{+2.4}_{-2.3}$ & $6.9^{+2.3}_{-2.2}$
           & $8.5^{+2.8}_{-3.5}$ & $5.8^{+2.4}_{-2.1}$ & $6.1^{+3.8}_{-2.9}$ \\
$|t_P|$ & $46.7^{+3.1}_{-3.4}$ & $43.1^{+3.2}_{-3.4}$ & $46.1^{+3.2}_{-3.4}$ 
        & $46.2^{+3.2}_{-3.4}$ & $43.3^{+3.2}_{-3.4}$ & $47.0^{+3.1}_{-3.4}$ \\
$|t_V|$ & $32.5^{+3.6}_{-3.9}$ & $30.3^{+3.4}_{-3.7}$ & $31.9^{+3.6}_{-4.0}$
        & $33.8^{+3.7}_{-4.1}$ & $29.6^{+3.4}_{-3.7}$ & $30.9^{+4.2}_{-4.7}$ \\
$\delta_P$ & $(184\pm12)^\circ$ & $(181\pm8)^\circ$ & $(36\pm13)^\circ$ 
 & $(199\pm14)^\circ$ & $(182\pm14)^\circ$ & $(16^{+20}_{-23})^\circ$ 
 \\
  $\delta_V$ & $(-87\pm11)^\circ$ & $(-18^{+7}_{-8})^\circ$ & 
  $(-100^{+10}_{-12})^\circ$
& $(-102^{+19}_{-14})^\circ$ & $(-20^{+9}_{-10})^\circ$ & 
$(-105^{+19}_{-26})^\circ$
\\
 $|C_P|$ & $6.0^{+4.3}_{-4.1}$ & $5.3^{+4.3}_{-4.0}$ & $5.7^{+4.4}_{-4.1}$
        & $6.8^{+4.5}_{-4.2}$ & $5.6^{+4.2}_{-4.0}$ & $6.0^{+4.4}_{-4.1}$ \\
$|C_V|$ & $16.8^{+5.5}_{-5.7}$ & $13.1^{+5.7}_{-6.0}$ & $15.0^{+5.6}_{-5.8}$
 & $17.3^{+5.6}_{-5.8}$ & $13.1^{+5.7}_{-6.0}$ & $14.9^{+5.6}_{-5.8}$ \\ \hline
Fit properties:   & & \\
$\chi^2/{\rm d.f.}$ & 23.9/24 & 25.5/24 & 22.1/24
                    & 20.6/22 & 20.5/22 & 18.9/22 \\
\% c.l.             &   47\%  &   38\%  &   57\%
                    &   55\%  &   55\%  &   65\% \\ \hline
Derived quantities: & & & & & & \\
$|p'_V/p'_P|$       & 1 (input) & 1 (input) & 1 (input)
           & $0.96^{+0.12}_{-0.10}$ & $1.15\pm0.07$ & $1.10\pm0.12$ \\
$-s'_P/p'_P$        & $0.03\pm0.04$ & $0.00\pm0.04$
                    & $0.02^{+0.04}_{-0.05}$ 
  & $0.05^{+0.06}_{-0.05}$ & $0.03\pm0.04$ & $0.04\pm0.05$ 
\\
$s'_V/p'_V$         & $0.19^{+0.08}_{-0.07}$ & $0.24^{+0.09}_{-0.08}$
                    & $0.21^{+0.08}_{-0.07}$
  & $0.27^{+0.13}_{-0.12}$ & $0.16^{+0.08}_{-0.06}$ & $0.17^{+0.14}_{-0.09}$ 
\\
$|t_V/t_P|$   & $0.70\pm0.10$ & $0.70\pm0.10$ & $0.69^{+0.11}_{-0.10}$
           & $0.73\pm0.10$ & $0.68\pm0.10$ & $0.66\pm0.11$ 
\\
Arg($t_V/t_P$) & $(-91\pm10)^\circ$ & $(-19\pm9)^\circ$
               & $(43^{+12}_{-11})^\circ$
               & $(-87\pm10)^\circ$ & $(-20\pm10)^\circ$
               & $(38^{+12}_{-11})^\circ$ \\  
$C_P/t_V$      & $0.19^{+0.17}_{-0.13}$ & $0.17^{+0.18}_{-0.14}$
               & $0.18^{+0.18}_{-0.14}$
  & $0.20^{+0.17}_{-0.13}$ & $0.19^{+0.19}_{-0.14}$ & $0.19^{+0.19}_{-0.14}$ 
\\
$C_V/t_P$     & $0.36^{+0.15}_{-0.13}$ & $0.30^{+0.16}_{-0.15}$ & 
$0.33^{+0.15}_{-0.14}$
              & $0.38^{+0.15}_{-0.14}$ & $0.30^{+0.16}_{-0.15}$ & 
              $0.32^{+0.14}_{-0.13}$ \\
$|p_P|=0.230|p'_P|$ & $7.4\pm0.3$ & $7.5\pm0.3$
                    & $7.4\pm0.3$
                  & $7.5\pm 0.4$ & $7.4\pm0.3$ & $7.4\pm0.4$ 
\\
$|p_V|=0.230|p'_V|$ & $7.4\pm0.3$ & $7.5\pm0.3$
                    & $7.4\pm0.3$
                    & $7.2^{+0.8}_{-0.7}$ & $8.5^{+0.5}_{-0.6}$ & 
                    $8.1\pm0.8$ \\
$|s_P|=0.230|s'_P|$ & $0.2\pm0.3$ & $0.0\pm0.3$
                    & $0.1\pm0.3$
                   & $0.4\pm0.4$ & $0.3\pm0.3$ & $0.3^{+0.4}_{-0.3}$ \\
$|s_V|=0.230|s'_V|$ & $1.4\pm0.5$ & $1.8^{+0.6}_{-0.5}$
                    & $1.6\pm0.5$
           & $2.0^{+0.6}_{-0.8}$ & $1.3^{+0.6}_{-0.5}$ & $1.4^{+0.9}_{-0.7}$ \\
$|t'_V|=0.281|t_V|$ & $9.1^{+1.0}_{-1.1}$ & $8.5\pm1.0$
                    & $9.0^{+1.0}_{-1.1}$
           & $9.5^{+1.0}_{-1.2}$ & $8.3\pm1.0$ & $8.7^{+1.2}_{-1.3}$ \\
$|t'_P|=0.240|t_P|$ & $11.2^{+0.7}_{-0.8}$ & $10.3\pm0.8$
                    & $11.0\pm0.8$ 
        & $11.1\pm0.8$ & $10.4\pm0.8$ & $11.3^{+0.7}_{-0.8}$ \\
$|C'_P|=0.230|C_P|$ & $1.4^{+1.0}_{-0.9}$ & $1.2^{+1.0}_{-0.9}$
                    & $1.3^{+1.0}_{-0.9}$
                    & $1.6\pm1.0$ & $1.3^{+1.0}_{-0.9}$ & 
                    $1.4^{+1.0}_{-0.9}$ \\
$|C'_V|=0.230|C_V|$ & $3.8\pm1.3$ & $3.0^{+1.3}_{-1.4}$ & $3.4\pm1.3$
                   & $4.0\pm1.3$ & $3.0^{+1.3}_{-1.4}$ & $3.4\pm1.3$ \\
\end{tabular}
\end{ruledtabular}
\leftline{$^a$EWP contribution only.}
\end{table*}

\begin{table*}
\caption{Comparison of predicted and experimental branching ratios
and $CP$ asymmetries $A_{CP}$ for some $\Delta S = 0$ decays of $B$ mesons.
References are given in Table \ref{tab:dS0data}.  Predictions are shown only
for the fit with $p'_V/p'_P$ complex.
\label{tab:pred0}}
\begin{ruledtabular}
\begin{tabular}{llrrrcrrrc}
 & & $|\leftarrow$~~~~~~ & \multicolumn{2}{c}{$\cB$ (units of $10^{-6}$)}
 & ~~~~~~$\rightarrow |$ & $|\leftarrow$~~~~~~ & \multicolumn{2}{c}{$A_{CP}$}
 & ~~~~~~$\rightarrow |$ \\
\multicolumn{2}{c}{Mode} &
\multicolumn{3}{c}{Prediction} & Expt.\ &
\multicolumn{3}{c}{Prediction} & Expt.\ \\
 & & $\gamma=26^\circ$ & $\gamma=63^\circ$ & $\gamma=162^\circ$ & &
     $\gamma=26^\circ$ & $\gamma=63^\circ$ & $\gamma=162^\circ$ & \\ \hline
$B^+ \to$
    & $\lbar{K}^{*0} K^+$
        & $0.51^{+0.06}_{-0.05}$ & $0.50\pm0.05$ & $0.49\pm0.05$ & $<5.3$ 
        & 0    &    0   &    0    &         \\
    & $K^{*+} \bar K^0$
        & $0.47^{+0.12}_{-0.09}$ & $0.66\pm0.08$ & $0.60\pm0.12$ &         
        &  0   &  0     &    0    &         \\
    & $\rho^0 \pi^+$
        & $8.7^{+0.3}_{-0.4}$ &  $9.0\pm0.4$ & $9.2^{+0.5}_{-0.6}$ & $9.1
\pm 1.1$ 
        & $-0.24\pm0.04$ & $-0.16\pm0.04$ & $-0.15\pm0.03$ & $-0.17 \pm
0.11$ \\
    & $\rho^+ \pi^0$
         & $10.1^{+1.6}_{-1.7}$ & $11.8^{+1.6}_{-1.8}$ &
$10.2^{+1.6}_{-1.7}$ & $11.0 \pm 2.7$ 
 & $0.03^{+0.08}_{-0.10}$ & $-0.01\pm0.06$ & $0.16^{+0.04}_{-0.05}$
 & $0.23 \pm 0.17$ \\
    & $\rho^+ \eta^{~a}$
       & $11.1^{+1.3}_{-1.5}$ & $9.5^{+1.3}_{-1.5}$ &
$10.7^{+1.4}_{-1.5}$ & $8.9 \pm 2.7$
        & $-0.05^{+0.03}_{-0.02}$ & $-0.01\pm0.07$ & $-0.01\pm0.03$ &
$0.06 \pm 0.29$ \\
    & $\rho^+ \eta'^{~a}$
 & $6.4^{+0.7}_{-0.8}$ & $4.9\pm0.7$ & $5.8^{+0.7}_{-0.8}$ & $13.3 \pm
4.5$
        & $-0.06^{+0.04}_{-0.03}$ & $-0.01\pm0.08$ & $-0.03\pm0.04$ &
\\
    & $\omega \pi^{+~b}$
    & $6.2\pm0.5$ & $5.9\pm0.4$ & $5.6\pm0.7$ & $5.9 \pm 1.1$
        & $0.01^{+0.04}_{-0.03}$ & $-0.03^{+0.10}_{-0.13}$ &
$-0.03^{+0.02}_{-0.06}$ & $0.10 \pm 0.21$ \\
    & $\phi \pi^{+~b}$
        & $0.001^{+0.005}_{-0.001}$ & $0.001^{+0.002}_{-0.001}$ &
$0.001^{+0.003}_{-0.001}$
& $ < 0.41$
        &  0     &  0   &   0     &           \\
\hline
$B^0 \to$
    & $\ol K^{*0} K^0$
        & $0.47\pm0.05$ & $0.46\pm0.04$ &   $0.46\pm0.05$  &
        &   0  &   0    &    0    & \\
    & $K^{*0} \ol K^0$
        & $0.44^{+0.11}_{-0.08}$ & $0.61\pm0.08$ &  $0.55\pm0.11$   &
        &  0   &    0   &    0    & \\
    & $\rho^- \pi^+$
        & $9.5\pm1.9$ & $10.0\pm2.0$ & $9.9^{+1.7}_{-1.8}$   &
$10.2 \pm 2.0$
 & $-0.19\pm0.04$ & $-0.13\pm0.06$ & $-0.13\pm0.03$ & $-0.54 \pm 0.19$ \\
    & $\rho^+ \pi^-$
         & $13.8\pm2.2$ & $14.0\pm2.2$ & $14.0\pm2.1$ & $13.8 \pm 2.2$
 & $-0.06^{+0.05}_{-0.04}$ & $-0.01\pm0.08$ & $0.04^{+0.04}_{-0.05}$
 & $-0.16 \pm 0.15$ \\
    & $\rho^0 \pi^0$
       & $0.63^{+0.35}_{-0.23}$  & $0.60^{+0.45}_{-0.32}$ &
$0.74^{+0.48}_{-0.34}$ & $<2.5$
        & $0.43^{+0.09}_{-0.16}$ & $0.09^{+0.33}_{-0.31}$ &
$0.21^{+0.09}_{-0.13}$ & \\
    & $\rho^0 \eta^{~a}$
        & $0.43^{+0.27}_{-0.20}$ & $0.08^{+0.17}_{-0.08}$ &
$0.13^{+0.17}_{-0.09}$ &
$<5.5$
        & $-0.20^{+0.11}_{-0.08}$ & $-0.09\pm0.46$ &
$-0.14^{+0.22}_{-0.14}$ & \\
    & $\rho^0 \eta'^{~a}$
        & $0.35^{+0.17}_{-0.14}$ & $0.07^{+0.10}_{-0.05}$ &
$0.13^{+0.11}_{-0.07}$ & $<12$
        & $-0.10^{+0.09}_{-0.05}$ & $0.08^{+0.36}_{-0.30}$ &
$-0.11^{+0.16}_{-0.07}$ & \\
    & $\omega \pi^{0~b}$
       & $0.36^{+0.36}_{-0.23}$ & $0.11^{+0.21}_{-0.06}$  &
$0.14^{+0.26}_{-0.12}$ &
$<1.9$
        & $0.35^{+0.26}_{-0.35}$ & $-0.26^{+0.71}_{-0.60}$ &
$0.41^{+0.56}_{-0.66}$ & \\
    & $\omega \eta^{~a,b}$
        & $0.30^{+0.26}_{-0.19}$ & $0.44^{+0.33}_{-0.25}$ &
$0.40^{+0.30}_{-0.22}$ &
$<12$
        & $-0.13^{+0.13}_{-0.09}$ & $-0.00\pm0.22$ &
$-0.02^{+0.08}_{-0.11}$ & \\
    & $\omega \eta'^{~a,b}$
        & $0.26^{+0.17}_{-0.13}$ & $0.27^{+0.18}_{-0.14}$ &
$0.27^{+0.18}_{-0.14}$ &
$<60$
        & $-0.23^{+0.13}_{-0.09}$ & $-0.04\pm0.22$ &
$-0.11^{+0.11}_{-0.10}$ & \\
    & $\phi \pi^{0~b}$
        & $0.001^{+0.002}_{-0.001}$ & $0.000^{+0.001}_{-0.000}$ &
$0.000^{+0.002}_{-0.000}$
& $<5$
        &    0   &   0   &   0    & \\
    & $\phi \eta^{~b}$
        & $<0.001$ & $<0.001$ & $<0.001$ & $<9$
        &    0   &   0   &   0    & \\ 
    & $\phi \eta'^{~b}$
        & $<0.001$ & $<0.001$ & $<0.001$ & $< 1.0$
        &    0   &   0   &   0    & \\
$S_{\rho \pi}$ & & & & &
 & $-0.19^{+0.06}_{-0.07}$ & $-0.18^{+0.21}_{-0.17}$ &
$-0.17^{+0.18}_{-0.14}$
 & $-0.13 \pm 0.18$ \\
$\Delta S_{\rho \pi}$ & & & & &
 & $0.22^{+0.18}_{-0.17}$ & $0.28^{+0.04}_{-0.05}$ &
$0.38^{+0.05}_{-0.08}$
 & $0.33 \pm 0.18$ \\
\end{tabular}
\end{ruledtabular}
\leftline{$^a$No $S_V$ contribution included.  $^b$No $S_P$ contribution
included.}
\end{table*}
%

\begin{table*}
\caption{Same as Table \ref{tab:pred0} for $|\Delta S| = 1$ decays of $B$
  mesons.
\label{tab:pred1}}
\begin{ruledtabular}
\begin{tabular}{llrrrcrrrc}
 & & $|\leftarrow$~~~~~~ & \multicolumn{2}{c}{$\cB$ (units of $10^{-6}$)}
 & ~~~~~~$\rightarrow |$ & $|\leftarrow$~~~~~~ & \multicolumn{2}{c}{$A_{CP}$}
 & ~~~~~~$\rightarrow |$ \\
\multicolumn{2}{c}{Mode} &
\multicolumn{3}{c}{Prediction} & Expt.\ &
\multicolumn{3}{c}{Prediction} & Expt.\ \\
 & & $\gamma=26^\circ$ & $\gamma=63^\circ$ & $\gamma=162^\circ$ & &
     $\gamma=26^\circ$ & $\gamma=63^\circ$ & $\gamma=162^\circ$ & \\ \hline
$B^+ \to$
    & $K^{*0} \pi^+$
   & $9.7^{+1.1}_{-0.9}$ & $9.5\pm0.9$ & $9.4\pm1.0$ &$9.0 \pm 1.4$
        &   0   &    0    &   0  & \\
    & $K^{*+} \pi^0$
  & $22.1^{+4.2}_{-5.1}$ & $15.0^{+3.3}_{-2.8}$ & $18.2^{+5.0}_{-4.1}$ & $
< 31$
 & $0.01\pm0.03$ & $0.01\pm0.05$ & $-0.04^{+0.03}_{-0.02}$ &      \\
    & $K^{*+} \eta^{~a}$
        & $25.1^{+2.0}_{-0.9}$ & $23.4^{+1.4}_{-1.2}$ &
$25.1^{+2.3}_{-0.9}$ & $25.9 \pm 3.4$
   & $-0.00\pm0.02$ & $0.00\pm0.02$ & $-0.04\pm0.02$ & $0.10 \pm 0.12$ \\
    & $K^{*+} \eta'^{~a}$
        & $2.2^{+2.1}_{-0.7}$ & $2.8^{+1.2}_{-0.3}$ & $2.4^{+3.1}_{-0.6}$
& $ < 12$
        & $0.26^{+0.04}_{-0.16}$ & $0.01\pm0.16$ &
$0.20^{+0.02}_{-0.14}$ & \\
    & $\rho^0 K^+$
          & $4.8^{+1.6}_{-0.3}$ & $4.4^{+0.8}_{-0.6}$  &
$4.5^{+1.5}_{-0.3}$
& $4.1\pm0.8$
  & $0.24\pm0.04$ & $0.21\pm0.10$ & $0.19^{+0.04}_{-0.05}$ & \\
    & $\rho^+ K^0$
    & $9.0^{+2.2}_{-1.6}$ & $12.6\pm1.6$ & $11.4^{+2.3}_{-2.2}$ & $ <48$
        &   0    &   0   &  0    & \\
    & $\omega K^{+~b}$
         & $5.3^{+1.5}_{-0.3}$ & $5.0^{+0.8}_{-0.4}$ & $5.1^{+1.4}_{-0.3}$
& $5.4 \pm 0.8$
 & $0.24\pm0.04$ & $0.19\pm0.08$ & $0.18^{+0.04}_{-0.05}$
 & $-0.003 \pm 0.122$ \\
    & $\phi K^{+~b}$
& $8.6\pm0.3$ & $8.7\pm0.4$ & $8.6^{+0.2}_{-0.3}$  & $9.0 \pm 0.9$
        &  0   &   0    &   0   & $0.030 \pm 0.072$ \\
\hline
$B^0 \to$
    & $K^{*+} \pi^-$
        & $15.3^{+1.3}_{-1.1}$ & $12.4\pm0.9$ &
$15.5^{+1.2}_{-1.1}$ & $15.3 \pm 3.8$
 & $0.06\pm0.04$ & $0.01\pm0.10$ & $-0.03^{+0.05}_{-0.04}$ & $0.26 \pm
0.35$
\\
    & $K^{*0} \pi^0$
        & $1.3^{+1.3}_{-0.7}$ & $0.8^{+1.0}_{-0.6}$ & $0.7^{+1.0}_{-0.3}$
& $< 3.5$
        & $-0.19^{+0.06}_{-0.05}$ & $-0.03^{+0.22}_{-0.29}$ &
$-0.03^{+0.13}_{-0.10}$ & \\
    & $K^{*0} \eta^{~a}$
        & $18.4^{+1.7}_{-0.7}$ & $19.1^{+1.2}_{-1.1}$ &
$18.5^{+1.9}_{-0.9}$ & $17.8 \pm 2.0$
        & $-0.001^{+0.008}_{-0.006}$ & $0.001\pm0.006$ &
$-0.012^{+0.004}_{-0.003}$ & $0.05 \pm 0.10$ \\
    & $K^{*0} \eta'^{~a}$
        & $2.8^{+2.1}_{-0.5}$ & $3.0^{+1.2}_{-0.3}$ & $2.9^{+2.8}_{-0.5}$
& $ < 6.4$
        & $0.05^{+0.02}_{-0.04}$ & $0.00\pm0.03$ & $0.04^{+0.02}_{-0.03}$
& \\
    & $\rho^- K^+$
        & $10.1^{+2.7}_{-0.9}$ & $10.0^{+1.4}_{-1.3}$ &
$9.9^{+2.5}_{-1.1}$ & $9.0 \pm 2.3$
  & $0.21^{+0.04}_{-0.03}$ & $0.16^{+0.06}_{-0.07}$ &
$0.15^{+0.03}_{-0.04}$ & $0.19
\pm 0.12$
\\
    & $\rho^0 K^0$
        & $5.3^{+1.8}_{-1.5}$ & $7.2^{+2.1}_{-1.9}$ & $6.4^{+1.6}_{-1.8}$
& $< 12.4$
        & $-0.04\pm0.03$ & $-0.02\pm0.01$ & $-0.02\pm0.02$ & \\
    & $\omega K^{0~b}$
        & $3.9^{+1.1}_{-0.6}$ & $5.3^{+0.8}_{-0.4}$ & $4.9^{+1.1}_{-0.8}$
& $5.2 \pm 1.1$
        & $0.04\pm0.03$ & $0.02\pm0.02$ & $0.02\pm0.02$ & \\
    & $\phi K^{0~b}$
  & $7.9^{+0.2}_{-0.3}$ & $8.1\pm0.3$ & $8.0^{+0.2}_{-0.3}$ & $7.8 \pm
1.1$
        &  0   &   0    &   0    & \\
\end{tabular}
\end{ruledtabular}
\leftline{$^a$No $S'_V$ contribution included.  $^b$No $S'_P$ contribution
included.}
\end{table*}

Within the range $38^\circ \le \gamma \le 80^\circ$ \cite{CKMf}, the present
fits specify $\gamma$ to within a few degrees at the $1 \sigma$ level.  Values
corresponding to $\Delta \chi^2 \le 1$ above the minimum for $p'_V/p'_P = -1$
and complex are $\gamma = (65\pm6)^\circ$ and $(63\pm6)^\circ$,
respectively.  Ranges for $\Delta \chi^2 \le 3.84$ above the minimum
(95\% c.l.\ limits) are 54$^\circ$--75$^\circ$ and 51$^\circ$--73$^\circ$ in
the two fits.

The local minima found at $\gamma \simeq (26,162)^\circ$ are associated with
larger $C'_V$ and $P'_{EW,V}$, leading to larger predicted branching ratios
$\cB(B^+ \to K^{*+} \pi^0) = (22.1^{+4.2}_{-5.1}, 18.2^{+5.0}_{-4.1}) \times
10^{-6}$ versus the $15.0^{+3.3}_{-2.8}$ prediction for $\gamma \simeq
63^\circ$.  These predictions do not conflict with the present CLEO
\cite{Jessop:2000bv} bound of $31 \times 10^{-6}$ and are not too far below it.
The branching ratio of this decay should be measurable very soon.  The solution
for $\gamma \simeq 162^\circ$ also predicts a larger $CP$ asymmetry for $B^+
\to \rho^+ \pi^0$, closer to the present central value which is slightly
disfavored in the fits with $\gamma \simeq 63^\circ$.  The fits with $\gamma
\simeq (26,162)^\circ$ predict larger branching ratios and $CP$ asymmetries for
the color-suppressed all-neutral decay modes $B^0 \to \rho^0(\eta,\eta')$ and
$B^0 \to \omega \pi^0$ than do the fits with $\gamma \simeq 63^\circ$.

Several observables provide the main contributions to $\chi^2$.  These are
summarized in Table \ref{tab:chisq}.  The large $CP$ asymmetry in $B^0 \to
\rho^- \pi^+$ provides the largest $\Delta \chi^2$ in all three cases.  The
only other contributions with $\Delta \chi^2 \ge 1$ occur for $\gamma =
162^\circ$, with $\Delta \chi^2[A_{CP}(B^+ \to K^{*+} \eta)] = 1.3$, and for
$\gamma = 26^\circ$, with $\Delta \chi^2[\cB(B^0 \to \omega K^0)] = 1.3$. 

\begin{table}
\caption{Observables providing $\chi^2 \ge 1.5$ in at least one of the
three fits with complex $p'_P/p'_V$.
\label{tab:chisq}}
\begin{ruledtabular}
\begin{tabular}{lccc}
Observable & $\gamma=26^\circ$ & $\gamma=63^\circ$ & $\gamma=162^\circ$ \\
\hline
$A_{CP}(B^+ \to \rho^+ \pi^0)$ & 1.4 & 2.0 & 0.2 \\
$\cB(B^+ \to \rho^+ \eta')$    & 2.4 & 3.5 & 2.8 \\
$A_{CP}(B^0 \to \rho^- \pi^+)$ & 3.7 & 4.9 & 5.0 \\
$A_{CP}(B^0 \to \rho^+ \pi^-)$ & 0.4 & 1.0 & 1.7 \\
$A_{CP}(B^+ \to \omega K^+)$   & 4.0 & 2.6 & 2.4 \\
$S_{\phi K_S}$                 & 1.6 & 1.6 & 1.6 \\ \hline
Sum:                         & 13.5 & 15.6 & 13.7 \\
\end{tabular}
\end{ruledtabular}
\end{table}

We comment on several predictions of the fits with $\gamma \simeq 63^\circ$
which appear to be of general nature, not depending on specific assumptions
about $p'_V/p'_P$ or symmetry breaking.
 
All fits predict $|t_V| < |t_P|$.  In a factorization
picture $t_V$ involves the production of a $\pi^\pm$ ($f_\pi = 130.7$ MeV) by
the weak current, whereas $t_P$ involves production of a $\rho^\pm$ ($f_\rho
= 208$ MeV).  This inequality is therefore not so surprising.  One has
$|t_V/t_P| \simeq 0.68$ while $f_\pi/f_\rho \simeq 0.63$, suggesting within
factorization that the $B \to \rho$ and $B \to \pi$ form factors are similar.
The values of $|t_V|$ we find are comparable to that of $|t|$, the tree
amplitude in $B \to PP$, which was found in \cite{Chiang:2003rb} to be $|t| =
27.1 \pm 3.9$ eV.  One also finds $|t_P/t| \simeq f_\rho/f_\pi$, as expected
from factorization when the $\rho$--$\pi$ mass difference is neglected.

All fits predict $p'_V \simeq - p'_P$, even if this equality is not enforced.
This is largely due to the need for constructive interference between
$p'_V$ and $-p'_P$ in the decays $B \to K^* \eta$, as proposed in Ref.\
\cite{HJLP}.  We find $|p'_P/p'| \simeq 0.7$, where $p'$ is the
strangeness-changing penguin amplitude in $B \to PP$ decays, with magnitude
\cite{Chiang:2003rb} $|p'| = 45.7 \pm 1.7$ eV determined primarily by the
decay $B^+ \to K^0 \pi^+$.

The ratio $-s'_P/p'_P$, corresponding to the electroweak penguin contribution
to $s'_P$, is found to be $0.03\pm0.04$ for the $\gamma \simeq 63^\circ$ fit,
somewhat smaller than expectations \cite{EWPp}.
The central value of the ratio $C_P/t_V$ is about 0.2.  These
amplitudes interfere constructively in such processes as $B^+ \to \rho^0 \pi^+$
and $B^+ \to \omega \pi^+$.  The corresponding ratio for $C_V/t_P$ is slightly
larger, about 0.3, helping to enhance the predicted rates for $B^+ \to \rho^+
(\pi^0,\eta,\eta')$.  A similar constructive interference between
$C$ and $T$ terms appears to account for an enhancement of the $B^+ \to
\pi^+ \pi^0$ decay (see, e.g., \cite{BBNS}).

Small strong phases (mod $\pi$) are favored, 
implying that the relative strong
phase between $t_V$ and $t_P$ is small.  This is consistent with the 
prediction of QCD factorization methods \cite{Beneke:2003}.

The pattern of strong and weak phases obtained here is such that there is
a small amount of constructive tree-penguin interference in the $CP$-averaged
branching ratios for $B^0 \to \rho^- \pi^+$ and $B^0 \to K^{*+} \pi^-$,
and a small amount of destructive interference in $B^0 \to \rho^+ \pi^-$ and
$B^0 \to \rho^- K^+$.  The preference of the fits for large values of $\gamma$
(small values of $\cos \gamma$), when final-state phases are small mod
$\pi$, is due in part to the fact that these interference effects are
relatively small.

We shall comment below on details
which depend on specific assumptions.  First we discuss some specific decay
modes for which predictions are fairly stable over the range of assumptions.

\subsection{$\Delta S = 0$ decays}

The decay $B^+ \to \overline{K}^{*0} K^+$ is predicted to be dominated by the
$p_P$ amplitude and thus to have zero $CP$ asymmetry.  Its branching ratio
is expected to be about $0.5 \times 10^{-6}$, considerably below present
upper limits.  Any deviations from these predictions could indicate the
importance of an annihilation amplitude or, equivalently, important
rescattering effects.  One expects a similar or slightly larger prediction for
$B^+ \to K^{*+} \ol K^0$.

A non-zero negative $CP$ asymmetry is predicted for $B^+ \to \rho^0 \pi^+$,
as a result of the interference of the amplitudes $t_V+c_P$ and $p_V-p_P$.
In our favored solution (with $\gamma \simeq 63^\circ$) this arises as
the result of a small but non-negligible relative final-state phase 
$\delta_V$.
The same phase contributes a $CP$ asymmetry of opposite sign to $B^+ \to
\rho^0 K^+$ and $B^+ \to \omega K^+$, as we shall see below.  It appears to
be generated in our fit by the appreciable $CP$ asymmetry in $B^0 \to
\rho^- \pi^+$, and also leads to a non-zero prediction for $\Delta S_{\rho
\pi}$.

The decay $B^+ \to \rho^+ \pi^0$ is expected to have a branching ratio of 
about $12 \times 10^{-6}$, consistent with the recently reported level
\cite{Ocariz:2003}.

The decay $B^+ \to \phi \pi^+$, dominated by an amplitude coming from the
electroweak penguin in $s_P$, is included for completeness. We do not expect it
to be observed any time soon.  The corresponding predicted rates for $B^0 \to
\phi(\pi^0,\eta,\eta')$ are factors of (2,3,6) smaller, respectively.

The branching ratio for the newly observed decay $B^+ \to \rho^+ \eta$
\cite{Aubert:2003ru} is well reproduced, in part because of the enhancement
associated with the constructive interference between $t_P$ and $C_V$.  The
large value of $C_V$ is driven by the attempt to fit an even larger
branching ratio for $B^+ \to \rho^+ \eta'$ reported in the same experiment.
The penguin contributions are expected to cancel if $p_V = - p_P$, so both
of these decays are expected to have zero or very small $CP$ asymmetries.
This stands in contrast to the large asymmetries expected
for $B^+ \to \pi^+ \eta$ and $B^+ \to \pi^+ \eta'$ \cite{Chiang:2003rb}.  
In $B^+ \to \rho^+ \eta^{(\prime)}$, the QCD penguin contributions
associated with the $u \bar u$ and $d \bar d$ components of the
$\eta^{(\prime)}$ nearly cancel one another if $p_V \simeq -p_P$,
while in $B^+ \to \pi^+ \eta^{(\prime)}$, these contributions
reinforce one another.

We predict $\cB(B^+ \to \rho^+ \eta')/\cB(B^+ \to \rho^+ \eta) \simeq 1/2$, 
whereas the observed ratio exceeds 1.  This may reflect a shortcoming of our 
description of the $\eta'$ wave function, which has been argued in Ref.\
\cite{Beneke:2002jn} to contain important symmetry-breaking effects.

We are unable to accommodate the large central value of the $CP$ asymmetry in
$B^0 \to \rho^- \pi^+$.  This will be true of any formalism which respects the
rate difference relation (\ref{eqn:rateV}), which is seen in Eq.\
(\ref{eqn:brV}) to be poorly obeyed by central values.

\subsection{$|\Delta S| = 1$ decays}

The decay $B^+ \to K^{*0} \pi^+$, dominated by $p'_P$, is the main source of
information on that amplitude.  It is expected to have zero $CP$ asymmetry.
Better measurement of its branching ratio would reduce the errors on $|p'_P|$.

Electroweak penguin contributions play an important role in the large predicted
value $\cB(B^+ \to K^{*+} \pi^0) \simeq 15 \times 10^{-6}$.  As mentioned,
one may expect detection of this mode in the near future, and it may help to
choose among various local $\chi^2$ minima in Fig.\ 1.

The predictions for $B^+ \to K^{*+} \eta$ and $B^+ \to K^{*+} \eta'$ 
include a small tree contribution whereas no such contribution is expected
for the corresponding decays $B^0 \to K^{*0} \eta$ and $B^0 \to K^{*0} \eta'$.
This leads us to expect a slight enhancement of ${\cal B}(B^+ \to K^{*+} \eta)$
with respect to ${\cal B}(B^0 \to K^{*0} \eta)$, as suggested by the data.
The successful prediction of the rates for $B \to K^* \eta^{(\prime)}$
is due in large part to the inclusion of the EWP contribution,
interfering constructively in each case with the main QCD penguin term.
Small $CP$ asymmetries are predicted in $B^+ \to K^{*+} \eta$ and
$B^0 \to K^{*0} \eta$.  The considerable suppression of the decays involving
$\eta'$ reflects the destructive interference of the $p'_P$ and $p'_V$
contributions \cite{HJLP}.  The predictions are presented in the absence of
a singlet penguin ($S'_V$) contribution, which would primarily affect the
$B \to K^* \eta'$ processes.  For quantitative predictions as a function of
$|S'_V/p'_P|$ see Ref.\ \cite{Chiang:2001ir}.

In the absence of electroweak penguin contributions one would expect the
$B^+ \to \rho^0 K^+$ and $B^+ \to \omega K^+$ decay rates to be equal
\cite{Lipkin:1998ew}.  In fact, $\cB(B^+ \to \omega K^+)/\cB(B^+ \to \rho^0
K^+)=1.3\pm0.3$, consistent with~1. This implies that additional electroweak
penguin contributions in the form of the $2s'_P$ term in $A(B^+ \to \omega
K^+)$ (see Table \ref{tab:dS1}) do not considerably affect the branching ratio
of this decay.  A small but non-negligible positive $CP$ asymmetry of about 0.2
is expected in both these processes through interference of the $p'_V$ and
$t'_V+c'_P$ terms.

The process $B^+ \to \rho^+ K^0$, as mentioned, is expected to be governed
solely by the $p'_V$ term.  Measurement of its branching ratio would provide
valued information on $|p'_V|$.  As noted, the only upper limit on the
branching ratio, ${\cal B}(B^+ \to \rho^+ K^0) < 48 \times 10^{-6}$, comes
from the CLEO Collaboration, so it should be improved (or the decay discovered)
 very soon.

The rate difference sum rule (\ref{eqn:rateP}) involving $B^0 \to K^{*+} \pi^-$
is seen in Eq.\ (\ref{eqn:brP}) to be satisfied, though with large errors.
We predict small values for $A_{CP}(B^0 \to K^{*+} \pi^-)$ and $A_{CP}(B^0
\to \rho^+ \pi^-)$.

We predict a branching ratio for $B^0 \to K^{*0} \pi^0$ of only about
$10^{-6}$, below the present experimental upper bound of $3.5 \times 10^{-6}$.
The electroweak penguin component of the $c'_V$ amplitude is responsible for
interfering destructively with the $p'_P$ component in this process.  We expect
a corresponding {\it enhancement} in $\cB(B^+ \to K^{*+} \pi^0)$, since the
relative signs of $c'_V$ and the dominant penguin amplitude $p'_P$ are opposite
in the two processes.  One expects the sum rule (analogous to one discussed
recently in \cite{comb03})
\bea
\cB(B^+ \to K^{*+} \pi^0) & + &
\frac{\tau_+}{\tau_0} \cB(B^0 \to K^{*0} \pi^0) = \nonumber \\
\half \left[ \cB(B^+ \to K^{*0} \pi^+) \right. & + & \left.
\frac{\tau_+}{\tau_0} \cB(B^0 \to K^{*+} \pi^-) \right]
\eea
to hold to first order in $|t'_P/p'_P|$ and $|c'_V/p'_P|$.  The right-hand side
is $(12.7 \pm 2.2) \times 10^{-6}$.
The sum rule is only approximately obeyed by the predicted branching ratios
since quadratic terms in $|t'_P/p'_P|$ and $|c'_V/p'_P|$ are non-negligible.

The branching ratio and $CP$ asymmetry for $B^0 \to \rho^- K^+$ are
reproduced satisfactorily.  Since
these quantities enter into the rate difference relation (\ref{eqn:rateV}),
which is poorly obeyed [see (\ref{eqn:brV})], one suspects that it is the
experimental $CP$ asymmetry $A_{CP}(B^0 \to \rho^- \pi^+) = -0.54 \pm 0.19$
which is slightly out of line, as mentioned earlier.

The decay $B^0 \to \rho^0 K^0$ is predicted to have a branching ratio of
about $7 \times 10^{-6}$, not far below its experimental upper limit of
$12 \times 10^{-6}$.  The $CP$ asymmetry is expected to be very small.

We already noted the comparison of $B^0 \to \omega K^0$ and $B \to \phi K$
amplitudes in Section \ref{sec:test} as one test for $p'_V = -p'_P$.
Zero $CP$ asymmetry is expected.
We predict $A_{\phi K_S} = 0$, consistent with observation, but, as mentioned
\cite{Chiang:2003jn}, are unable to account for $S_{\phi K_S} = -0.15 \pm
0.70~(S=2.11)$, predicting instead the value $\sin(2 \beta) = 0.736 \pm 0.049$.

\subsection{Processes sensitive to assumptions}

Most parameters of the fits appear to be relatively stable.  This stability is
due in part to the inclusion of $S_{\rho \pi}$ and $\Delta S_{\rho \pi}$,
which are the only quantities in which the interference between $t_V$ and $t_P$
is probed directly.  Small changes in relative strong phases occur when
we relax the assumption that $p'_V/p'_P = -1$.  The changes in predicted
branching ratios and $CP$ asymmetries appear to be so small that they will
not be detected in the near future.

The least stable aspect of the fits is associated with the amplitudes $C_V$
(color-suppressed tree) and $P'_{EW,V}$, contributing to $s'_V$ and $c'_V$.
The need for a $C_V$ amplitude is associated with the large branching ratios
for $B^+ \to \rho^+ (\pi^0,\eta,\eta')$, but we still cannot fit the large
branching ratio for the last process.  The $\chi^2$ minima at $\gamma \simeq
(26,162)^\circ$ are associated with larger values of $C_V$ and $P'_{EW,V}$,
which lead to the predictions $\cB(B^+ \to K^{*+} \pi^0) = (22.1^{+4.2}_{-5.1},
18.2^{+5.0}_{-4.1}) \times 10^{-6}$. 

We have assumed no SU(3) symmetry breaking in relating the $\Delta S = 0$
penguin amplitudes $|p_{P,V}|$ to the $|\Delta S| = 1$ amplitudes $|p'_{P,V}|$.
This assumption will be checked in the future when the appropriate $B \to
K^* \ol K$ or $B \to \ol K^* K$ decay rates are compared with the predictions
in Table VI.
The corresponding assumption $|P/P'| = |V_{cd}/V_{cs}| = 0.230$ is
close to being checked in $B^+ \to K^+ \ol K^0$ decays, where it entails the
prediction ${\cal B}(B^+ \to K^+ \ol K^0) = (0.75 \pm 0.11) \times 10^{-6}$
\cite{Chiang:2003rb}, to be compared with the experimental upper limit of $2.2
\times 10^{-6}$ \cite{Bona:2003}.

We have assumed nonet symmetry and a particular form of octet-singlet mixing
in describing the decays $B \to K^* \eta$.  When an independent measurement of
the $p'_V$ amplitude becomes available through the decay $B^+ \to \rho^+ K^0$,
this assumption will receive an independent check.

As mentioned earlier, we assumed that the strong phase of the electroweak
penguin contribution $P'_{EW,P}$ to $s'_P$ and $c'_P$ is the same as that of
$p'_P$, 
and that the $P'_{EW,V}$ contribution differs in phase by $180^{\circ}$
with respect to $p'_V$.
It may be necessary to relax these assumptions in future fits once more
data become available involving these contributions.

\section{U-SPIN RELATIONS \label{sec:uspin}}

In the previous sections we have employed the complete flavor $SU(3)$
symmetry group, neglecting small annihilation-type amplitudes. A best fit was 
performed in order to calculate magnitudes and phases of $SU(3)$ amplitudes.
In the present section we will rely only on U-spin \cite{MLL,Uspin}, an
important subgroup of $SU(3)$, introducing U-spin breaking in terms of ratios
of decay constants.  U-spin will be shown to imply two quadrangle relations
among $|\Delta S|=1$ amplitudes and two quadrangle relations among 
$\Delta S=0$ amplitudes. This exhausts all sixteen $B^+\to VP$ decays given
in Tables III and IV. Relations will also be presented among penguin amplitudes
in strangeness changing and strangeness conserving decays, and among tree
amplitudes in these decays. Such relations may be used with branching ratio
measurements to constrain tree amplitudes in $|\Delta S|=1$ decays and penguin
amplitudes in $\Delta S=0$ decays.  This could give an indication about
potential $CP$ asymmetries in certain modes.  Expressions and values for decay
amplitudes calculated in previous sections, where assumptions stronger than
U-spin and U-spin breaking were made, must obey the quadrangle relations as
well as these constraints.

The U-spin subgroup of $SU(3)$ is the same as the I-spin (isospin) except that
the doublets with $U = 1/2, U_3 = \pm 1/2$ are
\be
\label{eqn:qks}
{\rm Quarks:}~~\left[ \begin{array}{c} |\half~~\half \rangle \\
                                      |\half -\! \half \rangle \end{array}
\right] = \left[ \begin{array}{c} |d \rangle \\ |s \rangle \end{array}
\right]~~,
\ee
\be
{\rm Antiquarks:}~~\left[ \begin{array}{c} |\half~~\half \rangle \\
                                      |\half -\!\half \rangle \end{array}
\right] = \left[ \begin{array}{c} |\bar s \rangle \\ -\!| \bar d \rangle
\end{array} \right]~~.
\ee
$B^+$ is a U-spin singlet, while $\pi^+ (\rho^+)$ and $K^+ (K^{*+})$ 
belong to a U-spin doublet,
\be
|0~~0 \rangle = |B^+\rangle = |u\bar b\rangle~~,
\ee
\be\label{eqn:doublet}
\left[ \begin{array}{c} |\half~~\half \rangle \\
                                      |\half -\!\half \rangle \end{array}
\right] = \left[ \begin{array}{c} |u\bar s \rangle = |K^+~(K^{+*})\rangle\\ 
-\!| u\bar d \rangle = -|\pi^+~(\rho^+)\rangle
\end{array} \right]~~.
\ee

Nonstrange neutral mesons belong either to a U-spin triplet or a U-spin
singlet.  The U-spin triplet residing in the pseudoscalar meson octet is
\be\label{eqn:tripletP}
\left[ \begin{array}{c} |1~~1 \rangle \\ |1~~0 \rangle \\ |1 -\!1 \rangle
\end{array} \right] = \left[ \begin{array}{c}
|K^0 \rangle = |d \bar s \rangle \\
\frac{\sqrt{3}}{2} |\eta_8 \rangle - \frac{1}{2} |\pi^0 \rangle
= \frac{1}{\sqrt{2}} | s \bar s - d \bar d \rangle \\
-\! |\ol K^0 \rangle = -\! |s \bar d \rangle \end{array} \right]~~,
\ee
and the corresponding singlet is
\be\label{eqn:singletP}
|0~~0 \rangle = \frac{1}{2} |\eta_8 \rangle + \frac{\sqrt{3}}{2} |\pi^0 
\rangle = \frac{1}{\sqrt{6}} | s \bar s + d \bar d - 2u \bar u \rangle~~.
\ee 
In addition the $\eta_1$ is, of course, a U-spin singlet.  We take $\eta_8
\equiv (2 s \bar s - u \bar u - d \bar d)/\sqrt{6}$ and $\eta_1
\equiv (u \bar u + d \bar d + s \bar s)/\sqrt{3}$. The physical $\eta$ and 
$\eta'$ are mixtures of the octet and singlet,
\be
\label{eqn:etamix}
\eta = \frac{2\sqrt 2}{3} \eta_8 - \frac{1}{3} \eta_1~~,~~~
\eta' = \frac{2\sqrt 2}{3} \eta_1 + \frac{1}{3} \eta_8~~.
\ee

The U-spin triplet in the vector meson octet is
\be\label{eqn:tripletV}
\left[ \begin{array}{c} |1~~1 \rangle \\ |1~~0 \rangle \\ |1 -\!1 \rangle
\end{array} \right] = \left[ \begin{array}{c}
|K^{*0} \rangle = |d \bar s \rangle \\
\frac{1}{\s} |\phi \rangle - \frac{1}{2} |\rho^0\rangle - 
\frac{1}{2}|\omega\rangle
= \frac{1}{\sqrt{2}} | s \bar s - d \bar d \rangle \\
-\! |\ol K^{*0} \rangle = -\! |s \bar d \rangle \end{array} \right]~~,
\ee
and the corresponding singlet is
\be\label{eqn:singletV}
|0~~0 \rangle_8 = \frac{1}{\sx} |\phi \rangle + \frac{\sqrt{3}}{2} |\rho^0 
\rangle - \frac{1}{2\st}|\omega \rangle = \frac{1}{\sqrt{6}} | s \bar s + 
d \bar d - 2u \bar u \rangle~~.
\ee 
The SU(3) singlet vector meson is $|0~0 \rangle_1 =
(|\phi\rangle + \sqrt{2}|\omega\rangle)/\st$.

The $\Delta C=0,~\Delta S = 1$ effective Hamiltonian transforms like a $\bar s
\sim |\half \half \rangle$
component ($\Delta U_3 = \half$) of a U-spin doublet, while the $\Delta C=0,
\Delta S = 0$ Hamiltonian transforms like a
$\bar d \sim -|\half -\half \rangle$ component ($\Delta U_3 =
-\half$) of another U-spin doublet.  
Since the initial $B^+$ meson is a U-spin singlet, the final states are U-spin
doublets. The $VP$ states can be formed from a $K^{*+}$ or
a $\rho^+$ belonging to a U-spin doublet (\ref{eqn:doublet}), while the 
pseudoscalar meson  belongs to the two U-spin singlets, (\ref{eqn:singletP})
and $\eta_1$, and to the U-spin triplet (\ref{eqn:tripletP}). These $VP$ states
resemble the corresponding $PP$ states studied within U-spin in Ref.\
\cite{Chiang:2003rb}, the only 
difference being that the $K^+$ and $\pi^+$ are now replaced by $K^{*+}$ and
$\rho^+$. Alternatively, one may choose the pseudoscalar ($K^+$ or $\pi^+$) in
a U-spin doublet, while the vector meson resides in the two U-spin singlets 
(\ref{eqn:singletV}) and $|0~0 \rangle_1$ and in the triplet
(\ref{eqn:tripletV}).

One finds four cases in each of which four physical amplitudes are expressed in 
terms of three U-spin amplitudes, corresponding to final states in which one of 
the final mesons is a member of a U-spin doublet while the other belongs to the 
two U-spin singlets and the U-spin triplet. This implies two quadrangle
relations among $\Delta S=1$ amplitudes and two quadrangle relations among
$\Delta S=0$ amplitudes: 
\bea
\label{eqn:quadrangles}
2\s A(K^{*+}\eta) & + & A(K^{*+}\eta') \nonumber \\
& = & \sx A(\rho^+K^0) + \st A(K^{*+}\pi^0)~, \\
2\s A(\rho^+\eta) & + & A(\rho^+\eta') \nonumber \\
& = & \st A(\rho^+\pi^0) - \sx A(K^{*+} \bar K^0)~,\\
A(\rho^0 K^+) & + & A(\omega K^+) \nonumber \\
& = & \s A(\phi K^+) - \s A(K^{*0}\pi^+)~,\\
A(\rho^0\pi^+) & + & A(\omega \pi^+) \nonumber \\
& = &\s A(\phi\pi^+) + \s A(\bar K^{*0} K^+)~.
\eea
The first two relations are straightforward generalizations of corresponding
relations obtained for $B\to PP$ \cite{Chiang:2003rb}. The last quadrangle is
expected to
be squashed, since the two terms on the right-hand-side contain no tree 
amplitude and are expected to be smaller than each of the two terms on the 
left-hand-side. (See expressions and values in Table III.) All four relations 
among complex amplitudes hold separately for $B^+$ and $B^-$ decays. 

One may decompose the $\Delta S =1$ and $\Delta S=0$ effective Hamiltonians
into members of {\em the same} two U-spin doublets multiplying given 
CKM factors \cite{Uspin},  
\bea
\label{eqn:Hs}
{\cal H}_{\rm eff}^{\bar b\to\bar s} & = & V^*_{ub}V_{us}O^u_s + 
V^*_{cb}V_{cs}O^c_s~~,\\
\label{eqn:Hd}
{\cal H}_{\rm eff}^{\bar b\to\bar d} & = & V^*_{ub}V_{ud}O^u_d + 
V^*_{cb}V_{cd}O^c_d~~.
\eea
Hadronic matrix elements of the two U-spin doublet operators, $O_{d,s}^u$ and 
$O_{d,s}^c$, will be denoted $A^u$ and $A^c$ and will be referred to as tree 
and penguin amplitudes, where the latter include electroweak penguin
contributions.  Note that these amplitudes multiply different CKM factors in
$|\Delta S|=1$ and $\Delta S=0$ processes. The expressions (\ref{eqn:Hs}) and
(\ref{eqn:Hd}) imply relations among penguin amplitudes $A^c$ 
in strangeness changing and strangeness preserving processes and identical 
relations among corresponding tree amplitudes $A^u$. 

Starting with processes involving $\eta$ and $\eta'$, one may simply transcribe
results obtained for $B \to PP$ \cite{Chiang:2003rb}, replacing $\pi^+$ and
$K^+$ by $\rho^+$ and $K^{*+}$. Thus, one finds expressions for $\Delta S =0$
penguin amplitudes in terms of sums of two $|\Delta S| =1$ penguin amplitudes
which are expected to dominate these processes \cite{Chiang:2003rb}, 
\bea
\label{eqn:Aceta}
A^c(\rho^+ \eta) & = & 
A^c(K^{*+} \eta)  - \frac{2}{\st}A^c(\rho^+ K^0)~~,\\
\label{eqn:Aceta'}
A^c(\rho^+ \eta') & = &
A^c(K^{*+} \eta') -\frac{1}{\sx}A^c(\rho^+ K^0)~~.
\eea
Since all amplitudes involve unknown strong phases, these are in general
triangle relations. Assuming that the two penguin amplitudes on the right-hand
sides of each of Eqs.~(\ref{eqn:Aceta}) and (\ref{eqn:Aceta'}) dominate the
respective processes, the rates of these processes may be used to obtain
constraints on the penguin amplitudes on the left hand sides. For this purpose
one would need to measure $\cB(B^+\to \rho^+K^0)$ and improve the upper
bound on $\cB(B^+ \to K^{*+}\eta')$.

Similarly, one obtains expressions for $\Delta S =1$ tree amplitudes in 
terms of sums of two $\Delta S =0$ tree amplitudes,
\bea
\label{eqn:Aueta}
A^u(K^{*+} \eta) & = & 
A^u(\rho^+ \eta) + \frac{2}{\st}A^u(K^{*+} \ol K^0)~~,\\
\label{eqn:Aueta'}
A^u(K^{*+} \eta') & = & 
A^u(\rho^+ \eta') + \frac{1}{\sx}A^u(K^{*+} \ol K^0)~~.
\eea
The second terms on the right-hand-sides vanish in the approximation of
neglecting annihilation amplitudes. This provides two equalities between tree
amplitudes in $B^+\to K^{*+}\eta~(\eta')$ and $B^+ \to \rho^+ \eta~(\eta')$.
(In Tables III and IV these amplitudes are given by $t'_P$ and $t_P$,
respectively.)
The observed amplitude $A(B^+ \to \rho^+ \eta) = 31.2 \pm 4.7$ eV (see Table
\ref{tab:dS0}) then implies, via Eq.~(\ref{eqn:Aueta}), that the tree
contribution in $B^+ \to K^{*+} \eta$ is $(31.2 \pm 4.7) \times \bar\lambda
(f_{K^*}/f_\rho) \simeq (8 \pm 1)$ eV.
This is approximately the value calculated in Table IV.

Another set of U-spin relations, applying separately to penguin and tree
amplitudes, can be derived for decay amplitudes involving $\rho^0,~\omega$ and
$\phi$.  Physical amplitudes, consisting of penguin and tree contributions, may
be decomposed into U-spin amplitudes,
\bea
3A(\omega\pi^+) & = & -A^d_0 - 3A^d_1 + \s B^d_0~~,\\
3A(\phi \pi^+) & = & \s A^d_0 + 3\s A^d_1 + B^d_0~~,\\
3A(\omega K^+) & = & -A^s_0 + 3A^s_1 +\s B^s_0~~,\\
3A(\phi K^+) & = & \s A^s_0 -3\s A^s_1 + B^s_0~~,\\
A(\rho^0\pi^+) & = & A^d_0 - A^d_1~~,\\
A(\ol K^{*0} K^+) & = & - 2\s A^d_1~~,\\
A(\rho^0 K^+) & = & A^s_0 + A^s_1~~,\\
A(K^{*0}\pi^+) & = & -2\s A^s_1~~,
\eea
where $A_0,~A_1$ and $B_0$ correspond to final states with vector mesons in  
U-spin singlet and triplet in the octet and in the SU(3) singlet, respectively.
The superscripts $d$ and $s$ denote strangeness conserving and strangeness
changing amplitudes, respectively.

This decomposition implies several relations for penguin amplitudes,
\bea
\label{eqn:Acrho}
A^c(\rho^0\pi^+) & = & A^c(\rho^0 K^+) + A^c(K^{*0}\pi^+)/\s~~,\\
\label{eqn:Acomega}
A^c(\omega \pi^+) & = & A^c(\omega K^+) + A^c(K^{*0}\pi^+)/\s~~,\\
\label{eqn:Acphi}
A^c(\phi \pi^+) & = & A^c(\phi K^+) -  A^c(K^{*0}\pi^+)~~,\\
\label{eqn:AcK*K}
A^c(\ol K^{*0} K^+) & = & A^c(K^{*0}\pi^+)~~.
\eea

The penguin amplitudes on the right-hand-sides dominate the corresponding 
processes. Assuming $p'_P = - p'_V$, the two terms on the right-hand-side 
of Eq.~(\ref{eqn:Acrho}), $-p'_V/\s + p'_P/\s$ add up constructively,
while the two terms in Eq.~(\ref{eqn:Acomega}), $(p'_V + 2s'_P)/\s + p'_P/\s$,
add up destructively. Destructive interference occurs also in 
Eq.~(\ref{eqn:Acphi}), where the right-hand-side is $(p'_P + s'_P) - p'_P$.
Consequently, one expects large tree-penguin interference in $B^+ \to
\rho^0\pi^+$,
and small interference in decays to $\omega \pi^+$ and $\phi \pi^+$. 
No $CP$ asymmetries are expected in the two processes of Eq.~(\ref{eqn:AcK*K})
which are pure penguin in our approximation.

Similarly, one obtains U-spin relations for tree amplitudes,
\bea
\label{eqn:Aurho}
A^u(\rho^0 K^+) & = & A^u(\rho^0 \pi^+) - A^u(\ol K^{*0} K^+)/\s~~,\\
\label{eqn:Auomega}
A^u(\omega K^+) & = & A^u(\omega \pi^+) - A^u(\ol K^{*0} K^+)/\s~~,\\
\label{eqn:Auphi}
A^u(\phi K^+) & = & A^u(\phi \pi^+) +  A^u(\ol K^{*0} K^+)~~,\\
\label{eqn:AuK*pi}
A^u(K^{*0} \pi^+) & = & A^u(\ol K^{*0} K^+)~~.
\eea
In the approximation of neglecting annihilation amplitudes the second terms in
Eqs.~(\ref{eqn:Aurho})--(\ref{eqn:Auphi}) vanish. Thus, tree amplitudes within
each of the three pairs of $\Delta S =0$ and $|\Delta S|=1$ processes involving
a $K^+$ and a $\pi^+$ are equal.  Assuming that the tree amplitude dominates
$B^+ \to \omega \pi^+$, where the penguin amplitudes $p_P$ and $p_V$ interfere
destructively, Eq.~(\ref{eqn:Auomega}) implies a sizable tree amplitude in $B^+
\to \omega K^+$, $\approx 23~{\rm eV} \times \bar\lambda(f_K/f_\pi) \simeq 7$
 eV.
This value, calculated earlier in Table IV, permits a sizable tree-penguin
interference in this decay.  Table IV shows equal tree
amplitudes in $B^+\to \omega K^+$ and $B^+ \to \rho^0 K^+$.  This result is
beyond U-spin. All the tree amplitudes in Eqs.~(\ref{eqn:Auphi}) 
and (\ref{eqn:AuK*pi}) vanish in our approximation.
\medskip

\section{CONCLUSIONS \label{sec:concl}}

We have analyzed the decays of $B$ mesons to a charmless vector ($V$) and
pseudoscalar ($P$) meson in the framework of flavor SU(3).  The relative
magnitudes of tree and color-suppressed amplitudes extracted from data appear
consistent with the factorization hypothesis.  For example, the ratio of the
tree amplitude in which the current produces a vector meson to that in which
it produces a pseudoscalar is approximately $f_\rho/f_\pi$, and the ratio of
color-suppressed to tree amplitudes is approximately that in $B \to PP$ data.

Penguin amplitudes are also extracted from data.  Here we are not aware
of successful {\it a priori} predictions of their magnitudes.  For solutions
compatible with other determinations of $\gamma$ \cite{CKMf}, we find a
fairly stable pattern of small final-state phases (mod $\pi$), implying small
$CP$ asymmetries in all processes.
In particular, we do not expect a large $CP$ asymmetry in $B^0 \to \rho^-
\pi^+$.  We find a small relative strong phase between $t_V$ and $t_P$.
There exist also solutions for $\gamma$ outside the expected range; these
have larger final-state phases but cannot be excluded by present experiments.
Our preferred $\gamma \simeq 63^{\circ}$ fit 
favors a weak phase $\gamma$ within the range $57^\circ$--$69^\circ$
at the $1 \sigma$ level, and 51$^\circ$--73$^\circ$ at 95\% c.l.\ if one
restricts attention to the range $38^\circ$--$80^\circ$ allowed in fits to
other data \cite{CKMf}.

Predictions have been made for rates and $CP$ asymmetries in as-yet-unseen
decay modes.  Some of these modes, such as $B^+ \to \rho^+ K^0$,
$B^0 \to \rho^0 K^0$, and $B^+ \to K^{*+} \pi^0$ should be seen soon.  

A key assumption for which we have performed some tests and suggested others is
the relation \cite{HJLP} $p'_V/p'_P \simeq -1$ between penguin amplitudes in
which the spectator quark is incorporated into either a pseudoscalar meson or
a vector meson.  This relation is quite well satisfied, with the question of a
small relative strong phase between $p'_V$ and $-p'_P$ still open.

\section*{ACKNOWLEDGMENTS}

We thank M.~Beneke, P.~Chang, A.~H\"ocker, H.~Jawahery, A.~Kagan, H. Lipkin,
M.~Neubert, H.~Quinn,
J.~G.~Smith, W.~Sun, L.~Wolfenstein, F.~W\"urthwein, and H.~Yamamoto for 
helpful
discussions.  Part of this research was performed during the stay of M. G. and
J. L. R. at the Aspen Center for Physics.  This work was supported in part by
the United States Department of Energy, High
Energy Physics Division, through Grant Contract Nos.\ DE-FG02-90ER-40560 and
W-31-109-ENG-38.

\section*{APPENDIX:  $B \to \rho^\mp \pi^\pm$ RATES AND ASYMMETRIES}

In Table \ref{tab:dS0data} the $CP$-averaged branching ratio $\cB_{\rho \pi}
^{\pm \mp}$ quoted for the decay $B^0 \to \rho^\mp \pi^\pm$ is the sum of the
$CP$-averaged branching ratios for
$B^0 \to \rho^- \pi^+$ and $B^0 \to \rho^+ \pi^-$.  The $CP$ asymmetry $A_{CP}
(\rho^\mp \pi^\pm) = -0.14 \pm 0.08 \equiv A_{\rho \pi}$ is
\be
A_{\rho \pi} = \frac{\cB(\rho^+ \pi^-) - \cB(\rho^- \pi^+)}
{\cB(\rho^+ \pi^-) + \cB(\rho^- \pi^+)}~~,
\ee
where
\be
\cB(\rho^\pm \pi^\mp) \equiv \cB(B^0 \to \rho^\pm \pi^\mp) + \cB(\ol B^0 \to
\rho^\pm \pi^\mp)~~.
\ee
These quantities are related to the individual $CP$-averaged branching ratios
and $CP$ asymmetries by \cite{Hocker,Jawahery:2003}
$$
\frac{1}{2}[\cB(B^0 \to \rho^\pm \pi^\mp) + \cB(\ol B^0 \to \rho^\mp \pi^\pm)]
$$
\be
= \frac{1}{2}(1 \pm \Delta C \pm A_{\rho \pi} C) \cB_{\rho \pi}^{\pm \mp}~~, 
\ee
where $C=0.35 \pm 0.13 \pm 0.05$ and $\Delta C = 0.20 \pm 0.13 \pm 0.05$ are
measured in time-dependent decays \cite{Aubert:2003wr,Jawahery:2003}.  The
individual $CP$ asymmetries are
\bea
A_{CP}(B^0 \to \rho^- \pi^+) = \frac{A_{\rho \pi} - C - A_{\rho \pi} \Delta C}
{1 - \Delta C - A_{\rho \pi} C}~~, \\
A_{CP}(B^0 \to \rho^+ \pi^-) = - \frac{A_{\rho \pi} + C + A_{\rho \pi} 
\Delta C}
{1 + \Delta C + A_{\rho \pi} C}~~.
\eea
In calculating the entries in Table \ref{tab:dS0data} for the individual
branching ratios and asymmetries we have used the correlations among the
input variables \cite{Aubert:2003wr} to evaluate the experimental errors.

\end{document}